\documentclass[aps,nofootinbib,superscriptaddress,preprint]{revtex4-1}

\pdfoutput=1
\usepackage{amssymb, amsmath, graphicx}
\usepackage{color}
\usepackage{subfigure}
\usepackage[colorlinks=true, urlcolor=blue, linkcolor=blue,
citecolor=blue, hyperindex=true, linktocpage=true]{hyperref}

\newcommand{\non}{\nonumber \\}

\DeclareMathOperator{\sgn}{sgn}

   \newcommand{\ome}{\omega}

    \newcommand{\cD}{{\cal D}}

\newcommand{\cO}{{\cal O}}

\newcommand{\RR}{\mathbb{R}}

\newcommand{\re}{\ensuremath{\mathrm{Re}}}
\newcommand{\im}{\ensuremath{\mathrm{Im}}}
\newcommand{\e}{\ensuremath{\mathrm{e}}}

\newcommand{\pa}{\partial}

\newcommand{\rar}{\rightarrow}

\newcommand{\be}{\begin{equation}}
\newcommand{\ee}{\end{equation}}
\newcommand{\bea}{\begin{align}}
\newcommand{\eea}{\end{align}}

\newcommand\lrpar{\raise .8ex\hbox{$^\leftrightarrow$} \hspace{-9pt}
\partial}
\newcommand\lpar{\raise .8ex\hbox{$^\leftarrow$} \hspace{-9pt}
\partial}
\newcommand\rpar{\raise .8ex\hbox{$^\rightarrow$} \hspace{-9pt}
\partial}

\makeatletter
\def\blfootnote{\xdef\@thefnmark{}\@footnotetext}
\makeatother

\begin{document}
\title{Non-perturbative emergence of non-Fermi liquid behaviour in
  $d=2$ quantum critical metals}


\author{Balazs Meszena}
    \email{meszena@lorentz.leidenuniv.nl}
\affiliation{Institute Lorentz $\Delta$ITP, Leiden University, PO
  Box 9506, Leiden 2300 RA, The Netherlands}

\author{Petter S\"aterskog}
    \email{saterskog@lorentz.leidenuniv.nl}
\affiliation{Institute Lorentz $\Delta$ITP, Leiden University, PO
  Box 9506, Leiden 2300 RA, The Netherlands}

\author{Andrey Bagrov}
  \email{abagrov@science.ru.nl}
\affiliation{Institute for Molecules and Materials, Radboud University,
Heyendaalseweg 135, Nijmegen 6525 AJ, The Netherlands}
 
\author{Koenraad Schalm}
    \email{kschalm@lorentz.leidenuniv.nl}
\affiliation{Institute Lorentz $\Delta$ITP, Leiden University, PO
  Box 9506, Leiden 2300 RA, The Netherlands}
  
  \blfootnote{*, $\dagger$: These authors contributed equally to this work}

\begin{abstract}
We consider the planar local patch approximation of $d=2$ fermions at
finite density coupled to a critical boson. In the
quenched or Bloch-Nordsieck
approximation, where one takes the limit of fermion flavors $N_f\rar 0$, the fermion spectral function can be determined
 {exactly}. We show that one can obtain this non-perturbative answer
 thanks to a specific
identity of fermionic two-point functions in the planar
local patch approximation. The resulting spectrum is that of a
non-Fermi liquid: quasiparticles are not part of the 
exact fermionic excitation spectrum of the theory. Instead one finds 
continuous spectral weight with power law scaling 
excitations as in  a $d=1$ 
dimensional critical state. Moreover, 
at low energies there are three such excitations at three
  different Fermi surfaces, two with
a low energy Green's function $G \sim (\ome-v_*k)^{-1/2}$ and one with $G \sim |\ome+k|^{-1/3}$. 


\end{abstract}

\maketitle
\tableofcontents

\section{Introduction}

A complete classification of infrared universality classes for phases of quantum matter at finite density
is an open problem in condensed matter theory. Experimentally, a number of fermionic states of matter that exhibit breakdown of
the quasiparticle Fermi-liquid paradigm \cite{Landau} are known to exist,
e.g. the strange metallic phase of unconventional superconductors \cite{StrangeMetals} or the non-Fermi liquid phase of graphene \cite{Graphene1,Graphene2}. 
Theoretically, however, they are not understood. 
These phases are strongly interacting and this prevents the use of most conventional approaches that rely on perturbation  theory.

One important scenario which is widely believed to cause the partial destruction of Fermi surfaces
and substantial change of transport properties of the electronic state in high-$T_c$ compounds \cite{HighT}, heavy fermion systems \cite{HeavyFermions}, 
and Mott insulators \cite{SenthilMott, MisawaMott} is the interaction of electronic quasiparticles with gapless bosons. The underlying physics is the proximity of 
a quantum critical point and these bosons are the protected emergent gapless collective degrees of freedom \cite{Hertz, Sachdev1}. The nature of the fermion-boson interaction is determined by the precise details of the quantum critical point --- ferromagnetic \cite{Chubukov} or antiferromagnetic \cite{Chakravarty}
spin density waves, Kondo impurities \cite{Kondo}, etc. 

\begin{figure}
\begin{center}	(a){\raisebox{.34in}{\includegraphics[scale=0.25]{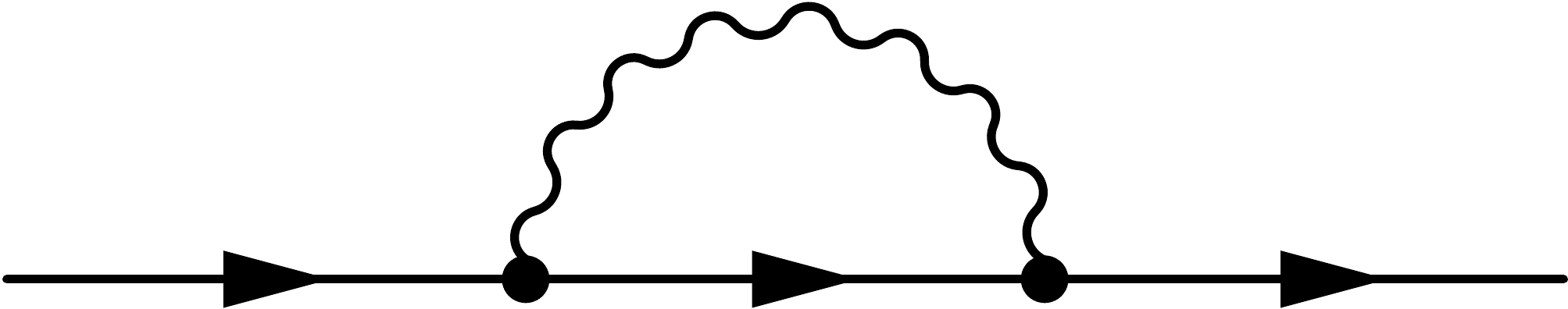}}~~~~~~~~}
(b)
{\raisebox{.15in}{\includegraphics[scale=0.25]{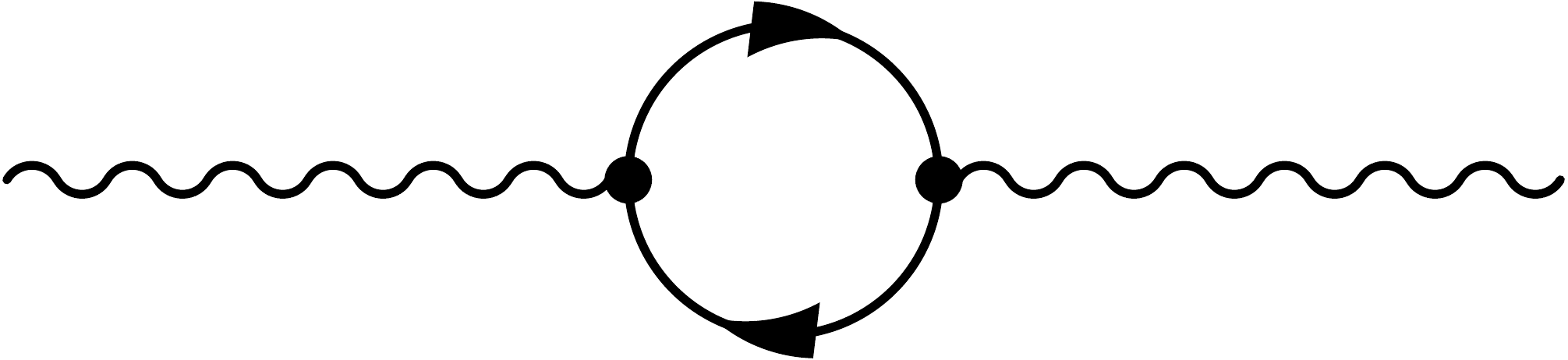}}}
\caption{
One loop corrections to fermion (a) and boson (b) self-energies due to Yukawa interaction. For a gapless boson (b) is the one loop contribution to Landau damping: this contribution to the self-energy can dominate in the IR. This term (b) is clearly proportional to the number of fermions $N_f$ in contrast to the one-loop correction to the fermion self-energy (a). In the limit $N_f \gg N_b$ the boson self-energy/Landau damping therefore dominates, whereas it is suppressed in the opposite limit $N_f \ll N_b$.}
\label{fig:one-loop}
\end{center}
\end{figure}

Qualitatively the simplest model 
that should already capture the non-trivial physics is the theory of spinless fermions at finite density interacting with a massless scalar through a 
straightforward Yukawa coupling, see e.g. \cite{Sachdev1}. The crucial physics that is thought to control the non-Fermi liquid behavior is the Landau damping: 
the quantum fermion-loop corrections to the boson two-point correlation function/self-energy. This is enhanced in a limit where the number of fermion flavors $N_f$ is much larger than the number 
of bosonic degrees of freedom ($N_f\gg N_b$);  this is easily seen at the one loop level where the diagram in Fig.~\ref{fig:one-loop}(b) is enhanced compared 
to Fig.~\ref{fig:one-loop}(a). In this regime the problem of a Fermi surface coupled to the Ising nematic and spin density wave 
 order parameters has been considered in \cite{Metlitski1,Metlitski2} and an extensive perturbative renormalization group analysis has been performed up to 
 three loops (higher order effects are investigated in \cite{Anomalous, Holder:2015pla}). However, as pointed out in these papers and \cite{SungSik1}, 
 in the (vector) large $N_f$ expansion one still needs to sum infinitely many diagrams. A well-defined expansion can be obtained by introducing an arbitrary dynamical
 critical exponent for the boson, $z_b$, as an extra control parameter \cite{Mross:2010rd}.

\begin{figure}
\begin{center}
\begin{align}
\nonumber
\underbrace{
\text{(a)}
\raisebox{.58in}{\includegraphics[scale=0.25]{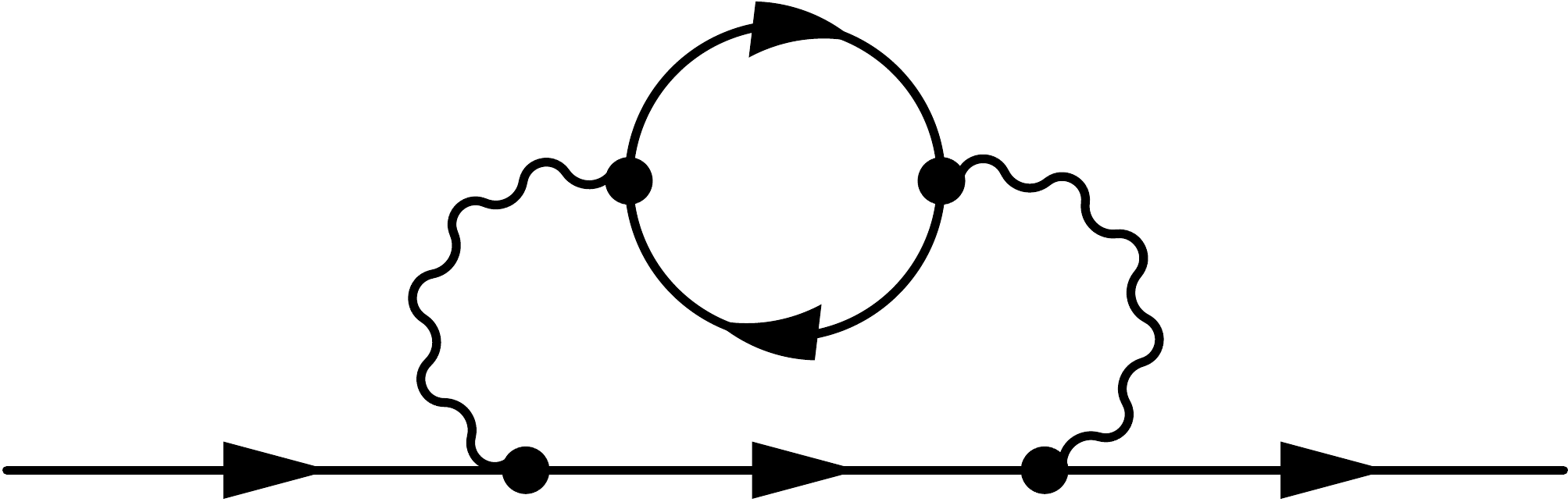}}
\underbrace{\text{(b)}
\raisebox{.2in}{\includegraphics[scale=0.25]{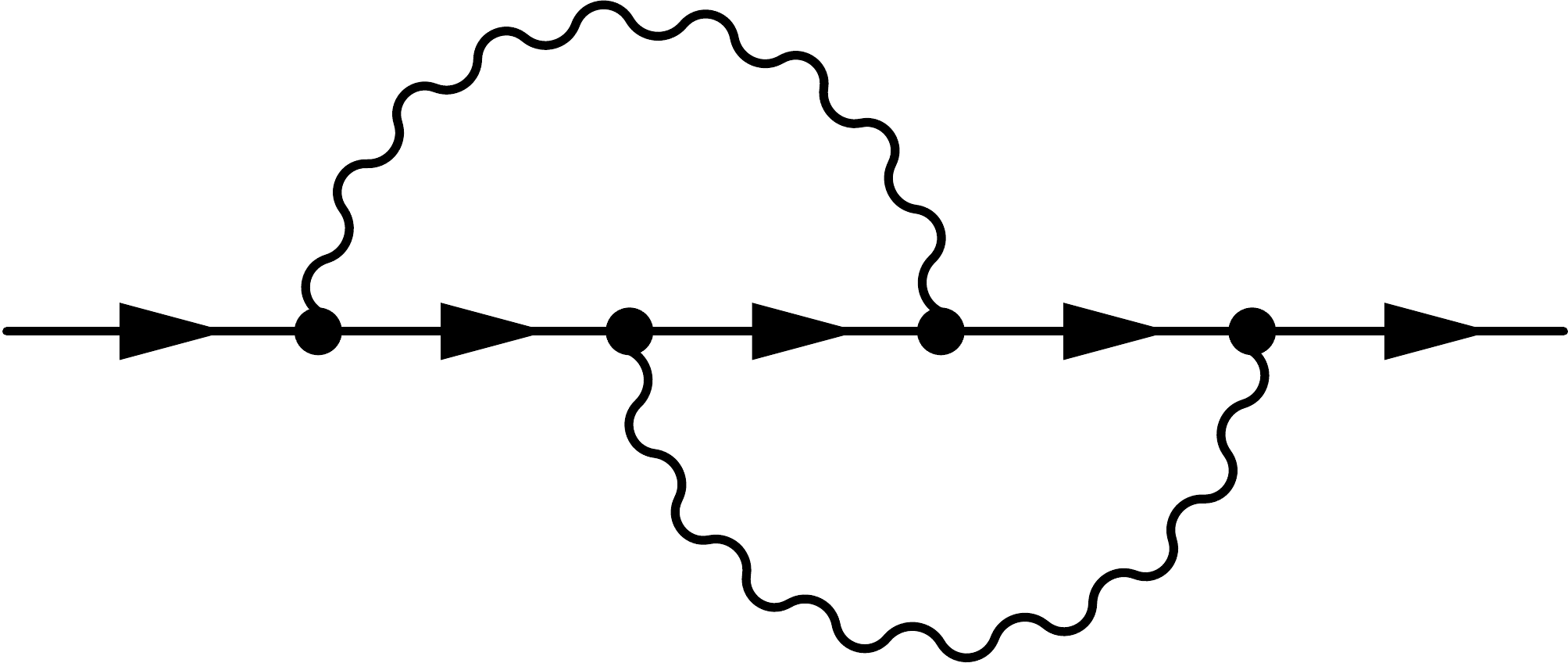}}
\underbrace{\text{(c)}
\raisebox{.58in}{\includegraphics[scale=0.25]{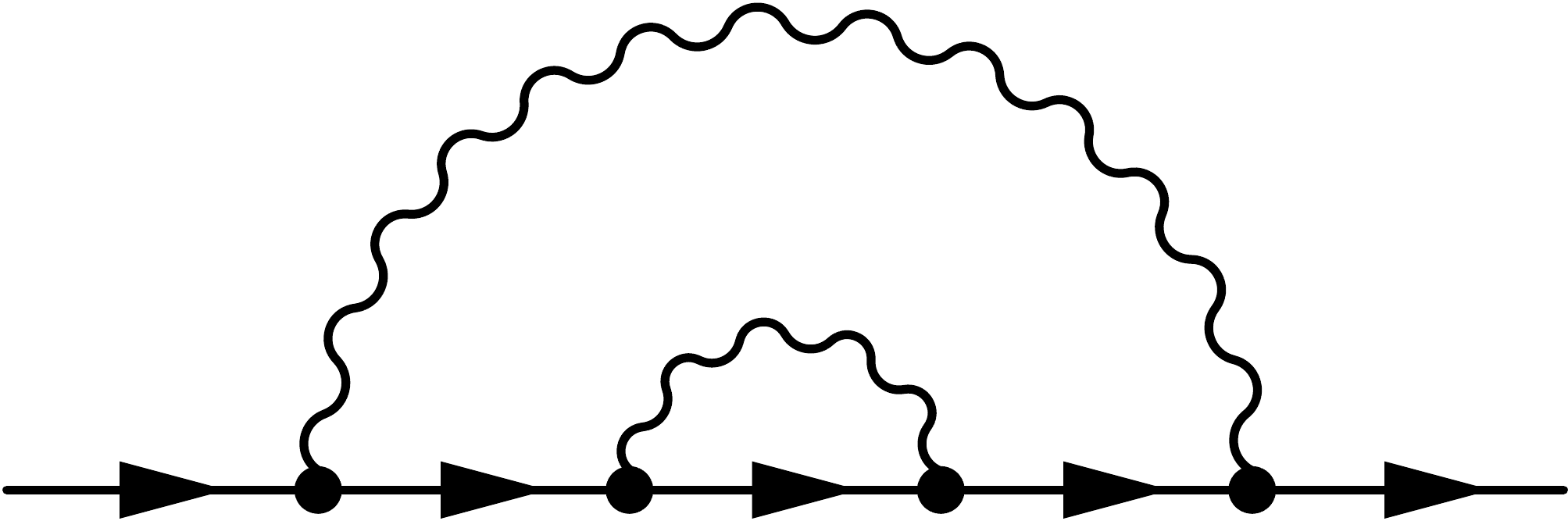}}
}_{\text{matrix large}~N}}_{\text{quenched approx.}}}_{\text{exact self-energy}}
\end{align}
\begin{align}
\nonumber
\raisebox{.38in}{\includegraphics[scale=0.35]{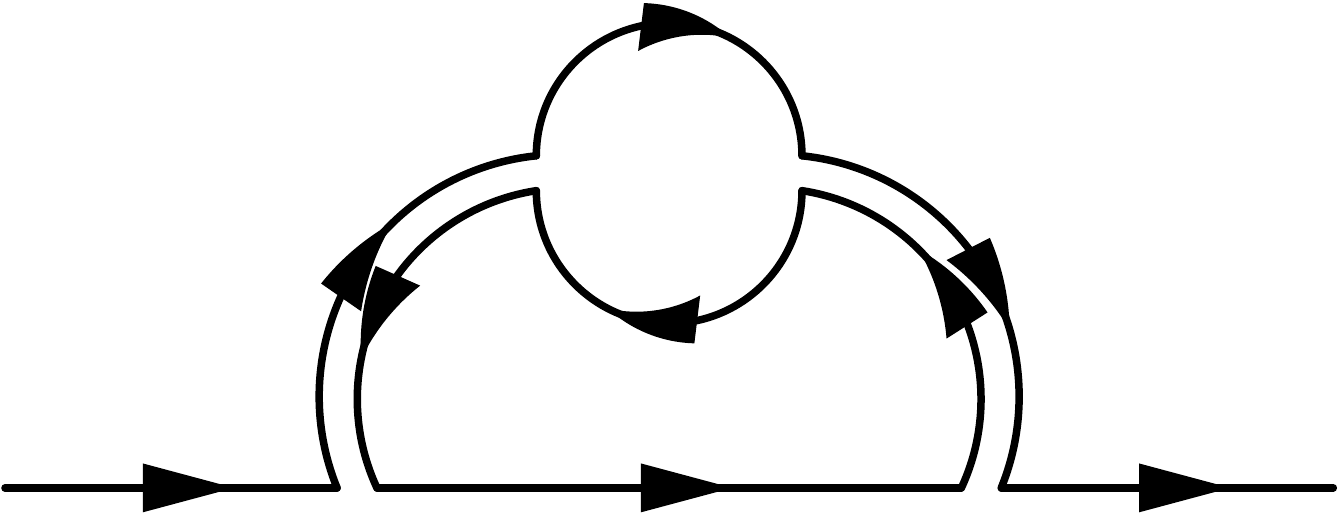}}~
~~~~~~\raisebox{-.02in}{\includegraphics[scale=0.35]{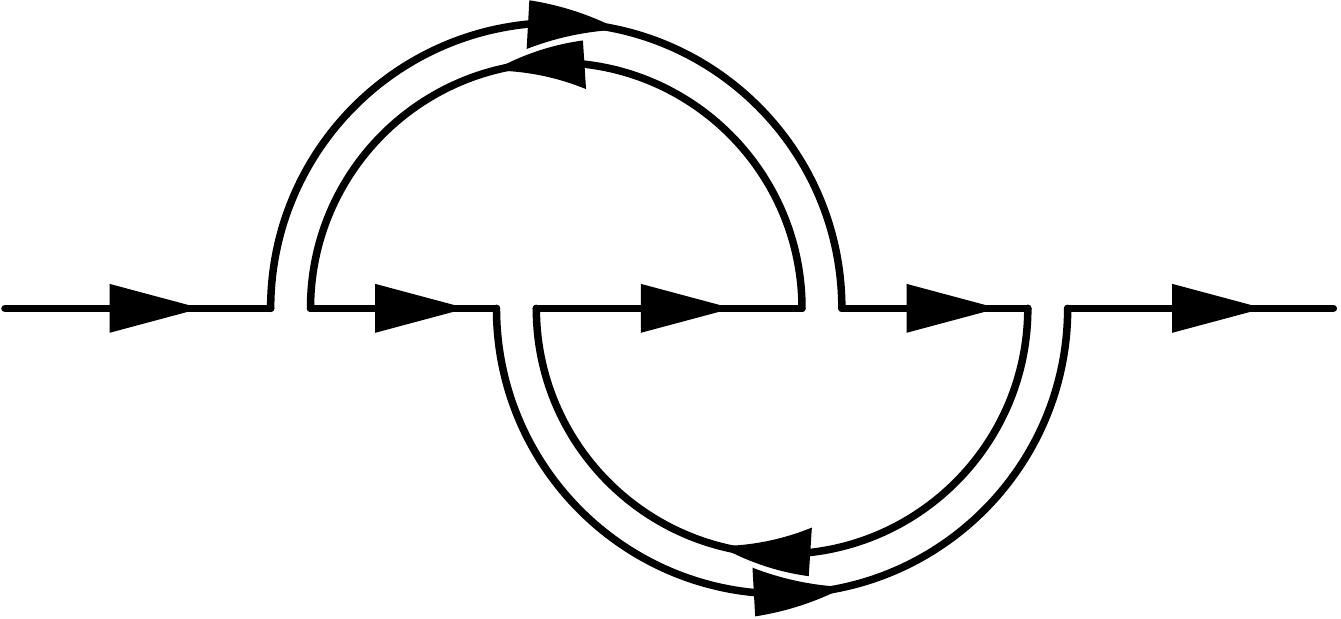}}
~~~~~~~\raisebox{.38in}{\includegraphics[scale=0.35]{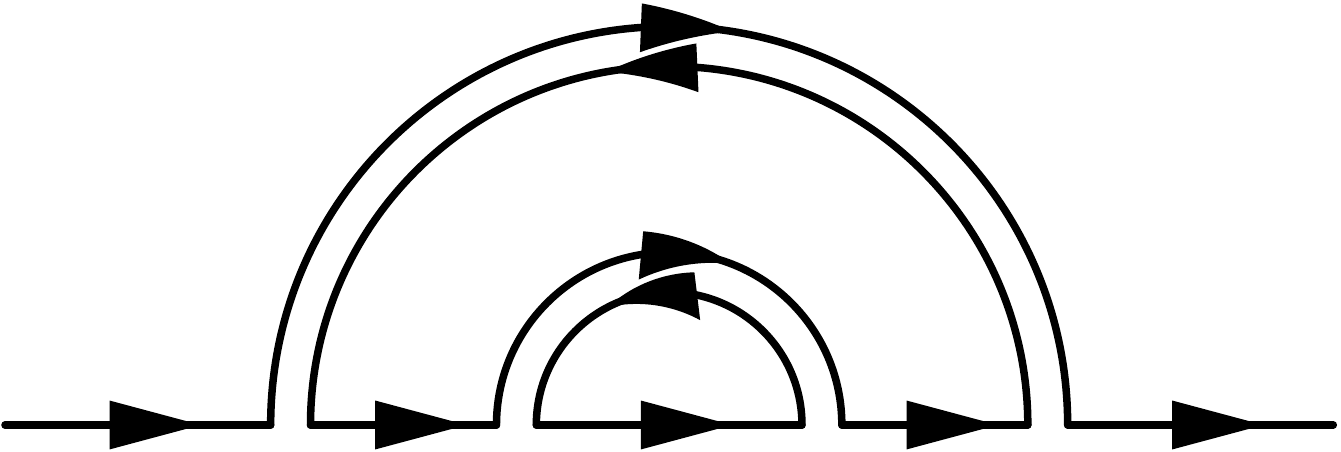}}
\end{align}
\caption{The two-loop contributions to the fermion self-energy in theory in Eq.~\eqref{eq:1}. The quenched limit where $N_f\rightarrow 0$ only suppresses the fermion loop contribution Fig. (a), whereas matrix large $N$ limits, where the boson scales as $N^2$ and the fermion as $N$ also suppress crossed diagrams of type (b). This is evident in double line notation.}
\label{fig:two-loop}
\end{center}
\end{figure}

Here, however, we show that Landau damping is not essential to obtain exotic non-Fermi liquid physics in the IR. We will study correlation functions of the theory 
in the opposite limit $N_f \rar 0;~N_b=1$. This quenched limit discards all fermion loop contributions. The strict quenched limit is more comprehensive than the recent matrix 
large $N$ expansions where the boson is taken to transform in the adjoint of an $SU(N)$ and the fermions in the fundamental, see e.g. the 
studies \cite{Stanford1, Stanford2, Stanford3, Stanford4}. In the matrix large $N$ limit one also has $N_b \gg N_f$, since $N_f=N$ and $N_b=N^2$. 
However in this case not only the diagrams with fermion loops but also diagrams with crossed boson lines are suppressed 
(Fig.~\ref{fig:two-loop}). By inspection of the associated momentum integral it is clear, however, that crossed boson corrections are important contributions to the IR physics. 
The IR of the quenched theory will therefore be different from the large $N$ matrix limit and perhaps closer to that of the full theory.

\begin{figure}[t!]
\centering
\begin{center}
\hspace*{-.1in}
\includegraphics[height=0.25\textheight]{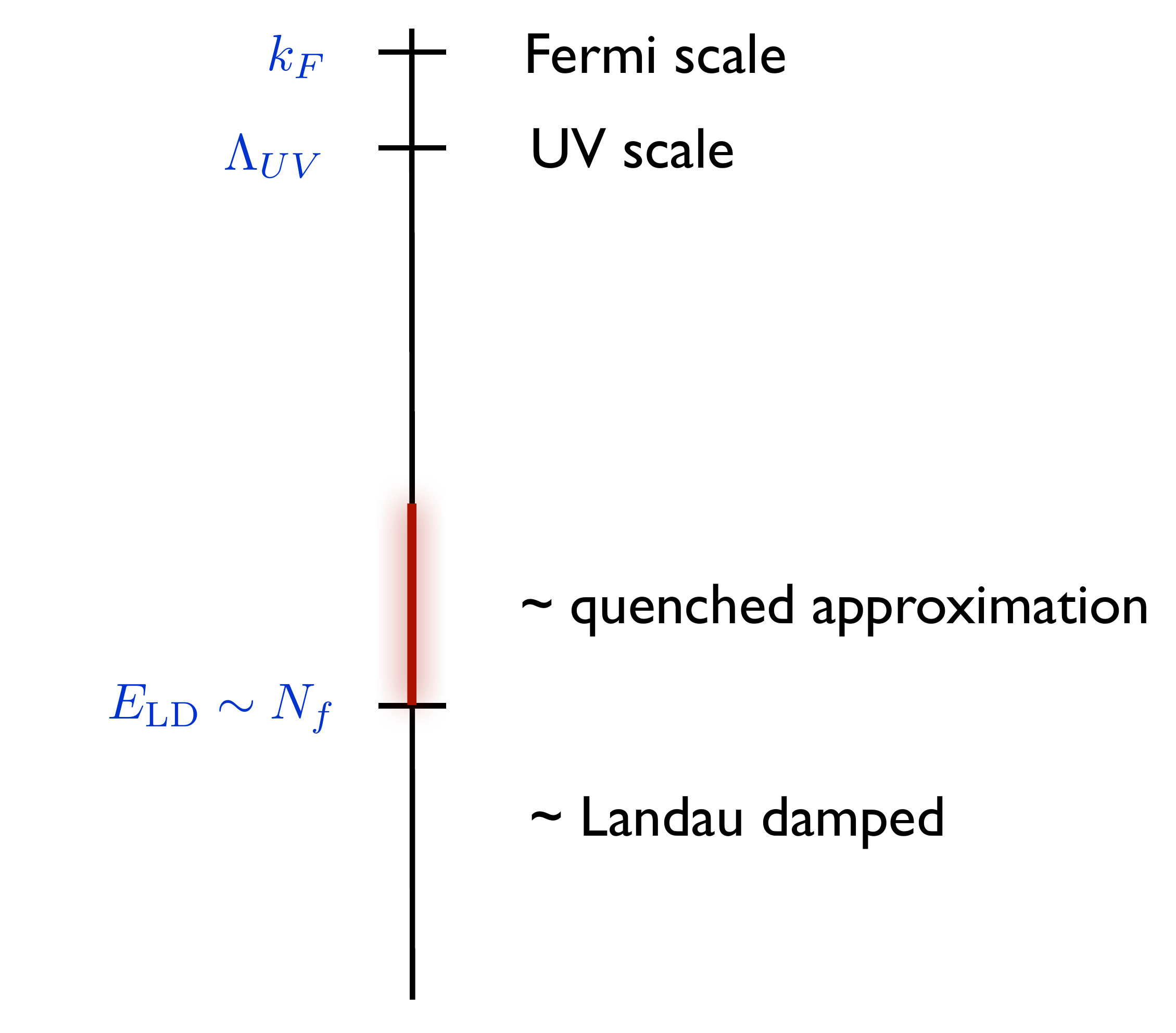}~~~~\raisebox{.2in}{\includegraphics[width=0.6\textwidth]{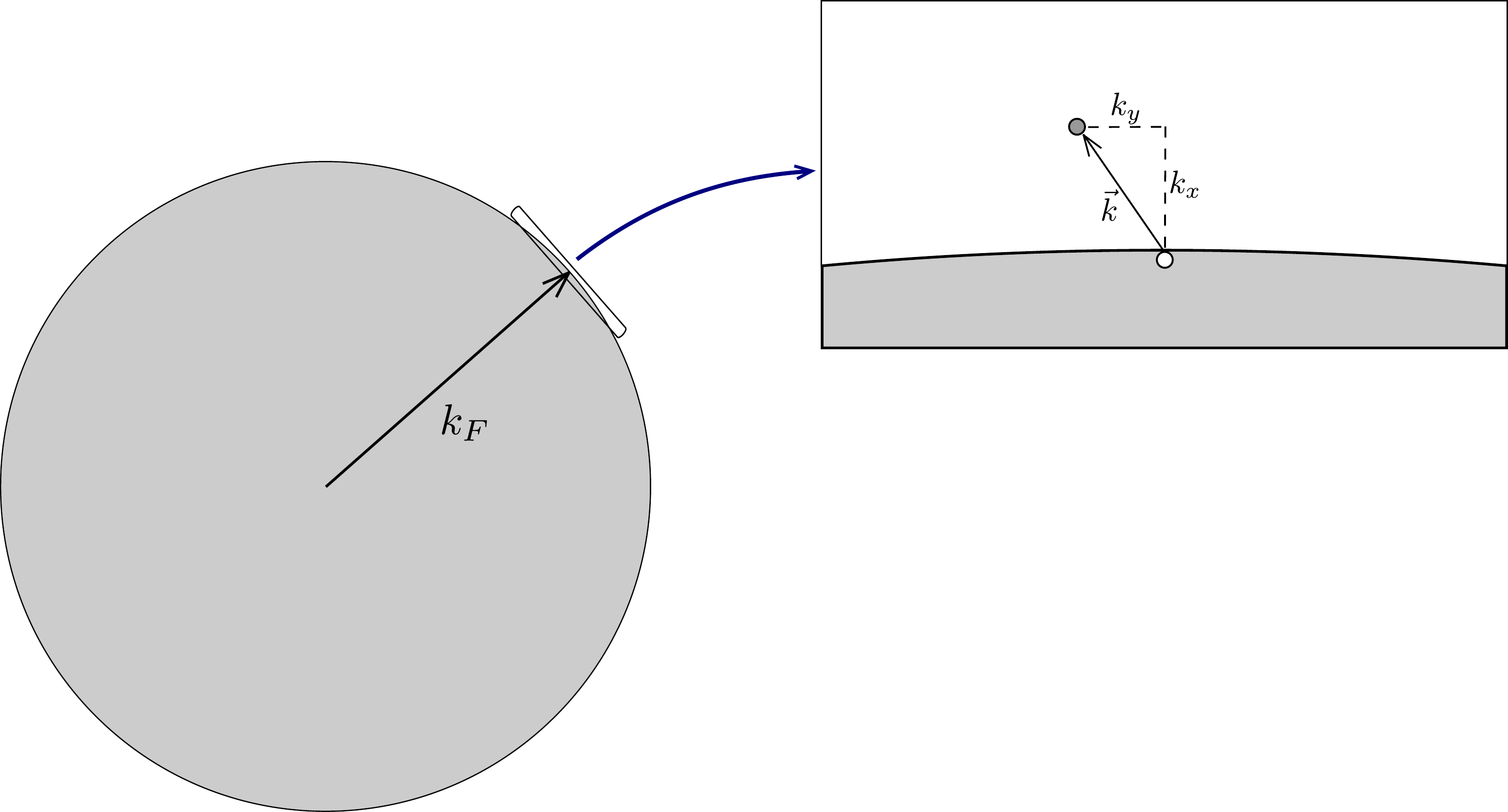}}
\end{center}
\caption{Left: The energy scales relevant to 2+1 dim. finite density fermions coupled to a massless boson. We assume the theory is already truncated at a UV cut-off $\Lambda_\mathrm{UV}$ 
lower than the Fermi scale $k_F$. The quenched $N_f \rightarrow 0$ limit focuses in on the intermediate energy regime right above the scale $E_{\mathrm{LD}} \sim N_f$ where 
Landau damping becomes important.
\\
Right: A small patch of the Fermi surface. In a small region near the Fermi level the surface curvature is negligible.
Fermionic excitations can acquire both orthogonal $k_x$ and tangent $k_y$ momenta, but the latter does not contribute to the kinetic energy of the excitation. 
In the $N_f\rightarrow 0$ limit each patch decouples from other parts of the Fermi surface.}
\label{fig:Patch}
  \label{fig:E-scales}
\end{figure}

Physically the quenched approximation means the following: as pointed out in, for example, \cite{Stanford4} there is a distinct energy scale where Landau damping becomes 
important. By considering small $N_f$ we are suppressing this scale and we are zooming in on the energy regime directly above the Landau damping scale 
$E_{\mathrm{LD}}$ (see Fig.~\ref{fig:E-scales}).\footnote{There are other ways to suppress the Landau damping physics, e.g. by considering large (UV) 
Fermi velocity, but we will not be considering these cases in this paper.}
We also make the assumption that that the cut-off of the boson is much smaller than the Fermi momentum ($k_F\gg\Lambda_\mathrm{UV}$). In this case the small $N_f$ limit also allows us 
to consistently focus on a small local patch  (see Fig.~\ref{fig:Patch})
around the Fermi surface where the fermionic excitations disperse linearly. The reason the curvature effects are negligible and the global structure of the Fermi surface 
becomes irrelevant, is that their influence on the IR physics is also through Landau damping. If the Landau damping is not negligible, one does have to either work with the 
full Fermi surface (i.e. in \cite{Stanford1, Stanford2, Stanford3, Stanford4} for the case of spherical Fermi surface) or at least consider the antipodal patch since the dominant 
contribution is coming from there \cite{Metlitski1,Metlitski2}.
Specifically, Landau damping depends on the Fermi-surface curvature $\kappa$ as $N_f/\kappa$. After the quenched approximation $N_f\rightarrow0$ for fixed $\kappa$, we may subsequently take $\kappa$ small as well.

The remarkable fact is that with these approximations the fermion Green's function can be determined exactly (directly in $d=2$ spatial dimensions). 
We achieve this by solving the differential equation for the Green's function in a background scalar field and then evaluating the bosonic path integral. 
Similar functional techniques have been used in high energy physics, for example in the study of high temperature QED plasma \cite{BN2}, lattice QCD \cite{Golterman:1994mk} or 
for solving the so-called Bloch-Nordsieck model (which is QED in the quenched approximation) \cite{BN1, BN3, BN4, Karanikas:1992aj}. In condensed matter context, 
the fact that the fermion spectral function is exactly solvable in these limits was also observed for finite density fermions coupled to a transverse gauge field by 
Khveshchenko and Stamp \cite{Khveshchenko1} and independently by Ioffe, Lidsky, Altshuler and Millis \cite{Altshuler1,Altshuler2}, though the latter solve the model by bosonization.

At the technical level, the reason the spectral function can be solved exactly in the quenched limit is that propagators of linearly dispersing fermions 
(the local patch approximation) obey special identities. These allow a rewriting of the loop expansion in such a way that it can be resummed completely, 
or rather that it can be recast as the solution to a tractable differential equation. We show this in section~\ref{sec:approximation}.
Note, that our method does not rely on renormalization group techniques. When we are to define an RG flow, we have to choose a proper decimation scheme. 
In relativistic field theories, it is natural to define the cut-off in a way that maintains the Lorentz invariance, while for the 
non-relativistic model of critical metals the choice of the cut-off is ambiguous, see e.g. \cite{Stanford1}.

These exact results in quenched approximation then allow us to establish that the IR fermion physics, even in the absence of Landau damping, 
is already that of a non-Fermi liquid. Specifically we show in section~\ref{sec:results} that:
\begin{itemize}
\item The naive free Fermi surface breaks apart into three.  A thin external shell of it splits apart from the rest, and we effectively 
 have three nested singular surfaces (see Fig. \ref{fig:instability}). This immediately follows from the fact that in a region around the original (free) 
 Fermi level the dispersion of fermion changes sign, $\frac{d\omega}{d k_x}<0$.
 This can be interpreted as a topological instability of the Fermi surface \cite{Anomalous}, as the dispersion curve must cross the Fermi energy two more times to 
 connect to the free UV theory. Luttinger's theorem nevertheless continues to hold.
\begin{figure}[t]
\centering
\includegraphics[width=0.5\textwidth]{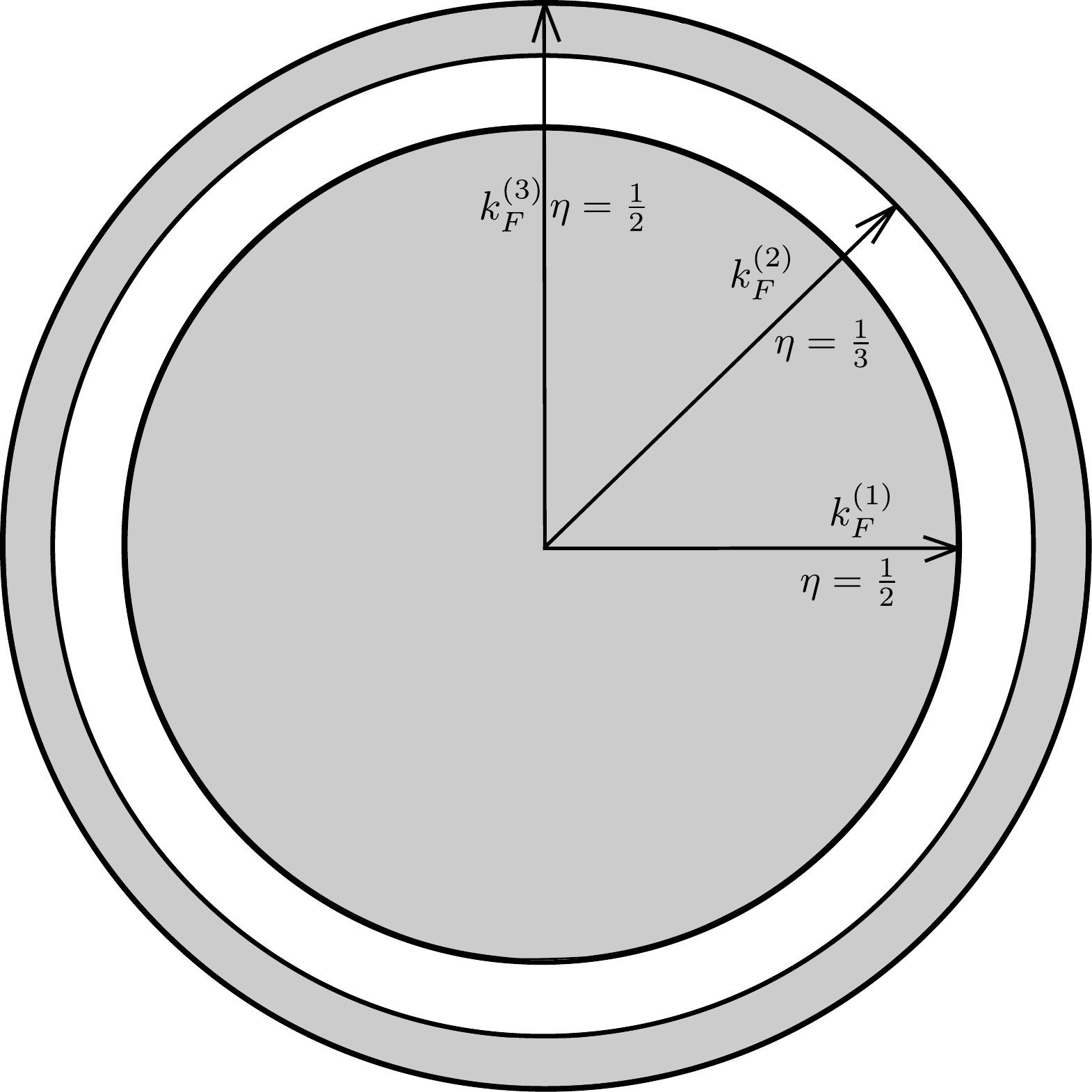}
\caption{
The emergent Fermi surface structure at low energy in the 2+1 dimensional quenched planar patch quantum critical metal. Due to the interactions the naive single Fermi surface is topologically unstable. The excitations around the Fermi surfaces are not well-defined quasiparticles, instead they have a continuous spectrum corresponding to a Green's function of the type $G(\omega,k) \sim (\omega-vk)^{-\eta}$ with scaling dimensions $\eta$.
The different values of $\eta$ at each of the emergent singular surfaces are mentioned.
}
\label{fig:instability}
\end{figure}
\item The Fermi-liquid quasiparticle pole is destroyed by the interaction with the critical boson.
Instead the spectrum is singular {\em everywhere} on the dispersion curve. 
Specifically near the three Fermi surfaces the singular Green's function takes a scaling form with different scaling dimensions. Around the original Fermi momentum 
the Green's function behaves as $G(\omega,k_x) \sim (\omega+ck_x)^{-1/3}$, where $c$ is the dispersion velocity of the boson; around the two split-off Fermi 
surfaces behaves as $G(\omega, k) \sim (\omega-v^{\ast}k_x)^{-1/2}$ with $v^{\ast}$ an emergent dispersion velocity $0<v^{\ast}<v_F$.
\end{itemize} 

We conclude with a brief outlook in section \ref{sec:conclusions}.


\section{2+1 dimensional quantum critical metals in the patch
  approximation}
 \label{sec:approximation}

As stated, the theory we study is that of $N_f$ spinless fermionic flavours 
at finite density minimally coupled to a critical (relativistically
dispersing) boson in $2+1$
dimensions. It has the Euclideanized action
\begin{align}
  S = \int \mathrm{d}x\mathrm{d}y\mathrm{d}\tau \left[ \psi^{\dagger}_j (-\partial_{\tau} -
    \frac{\nabla^2}{2m} +\mu) \psi^j
    +\frac{1}{2}\left(\partial_{\tau}\phi\right)^{2}+\frac{1}{2}\left(\nabla\phi\right)^{2}
    + \lambda \phi\psi^{\dagger}_j\psi^j \right]
\end{align}
with $j=1\ldots N_f$; we will show below that the rotation back to
real time has no ambiguities. Assuming that the theory is still weakly coupled at scales
much below the Fermi momentum $k_F$, we may make a local  
approximation around a patch of the Fermi surface and truncate the
fermion kinetic term to (Fig.~\ref{fig:Patch}) \cite{Sachdev1}
\begin{align}
\label{eq:1}
  S_{P}= \int \mathrm{d}x\mathrm{d}y \mathrm{d}\tau \left[ \psi^{\dagger}_j (-\partial_{\tau} +iv\partial_x) \psi^j
    +\frac{1}{2}\left(\partial_{\tau}\phi\right)^{2}+\frac{1}{2}\left(\partial_x\phi\right)^{2}+\frac{1}{2}\left(\partial_y\phi\right)^{2}
    + \lambda \phi\psi^{\dagger}_j\psi^j \right].
\end{align}
Two comments are in order. (1)
Though it is very well known that the leading ``Fermi surface
curvature'' correction to the kinetic term
$\mathcal{L}_{\mathrm{curv}}= -  \kappa \psi^{\dagger}_j \partial_y^2 \psi^j/2+...$
is a dangerously irrelevant operator important for fermion loops even at low energies, in the $N_f\rightarrow0$ limit (where there are no fermion loops) this operator is safely irrelevant and can be consistently neglected for physics below the scale set by $1/\kappa$. We will show here through exact results that the minimal theory in
Eq.~\eqref{eq:1} already has a very non-trivial IR. We shall comment
on the relevance of $\mathcal{L}_{\mathrm{curv}}$ to our results below.
(2) From a Wilsonian point of view, self-interactions of the boson
should also be included. We leave the effect of this term for future investigations and take
the action $S_{P}$ as given from here on and study it on its own.\footnote{Note that, 
although the kinetic term is effectively $(1+1)$-dimensional,
the properties of fermionic field are still strongly dependent on the dimensionality
of the system, because the fermions interact with the
$(d+1)$-dimensional boson. An instructive way to think about the
fermion dynamics in dimensions parallel to the Fermi surface, is to
Fourier transform in those directions. Because the kinetic term does
not depend on these directions, the parallel momenta act as additional
global quantum numbers. E.g. in $d=2$, one therefore has an
infinite set of one-dimensional fermionic subsystems, labeled by $k_y$. The Yukawa
interaction with the bosons then describes the interactions between
these many one-dimensional subsystems.}

Throughout the paper we are mostly interested in the case where the characteristic speed $c=1$ 
of the critical bosonic excitations
is larger than the Fermi velocity, $c>v$. This need not be the case in the UV. However, as was recently argued \cite{Stanford2,Stanford4}, the Fermi velocity decreases substantially under the RG flow
and because in our analysis we consider energies below a cut-off $\Lambda_\mathrm{UV}\ll k_\mathrm{f}$,
we take this condition for granted as a starting point.

In $d=2$ spatial dimensions the Yukawa coupling is relevant ---
  $\lambda$ has scaling dimension 1/2 --- and the theory
  will flow to
  a new IR fixed point.
Rather than focusing on a complete understanding of the IR of the
action Eq. \eqref{eq:1}, we will
focus only on understanding a single correlation function: the fermion
spectral function. Coupling the fermionic fields to external sources 
\begin{equation}
Z[J,J^{\dagger}]=\int\cD\psi^{j}\cD\psi_{j}^{\dagger}\cD\phi\exp\left(-S_{P}-J_{j}^{\dagger}\psi^{j}-\psi_{j}^{\dagger}J^j\right),
\end{equation}
the fermionic integral is Gaussian and can be easily evaluated yielding
\begin{align}
Z[J,J^{\dagger}]&=\int
\cD\phi\left(\det(G{}^{-1}[\phi])\right)^{N_f}\exp\left(-S_{b}[\phi]-\int\!
  d^{3}zd^3z'
  J_{i}^{\dagger}(z)G^{i}_{~j}[\phi](z;z')J^{j}(z')\right), \non
&=\int
\cD\phi\,\exp\left(-S_{b}[\phi]-S_{\text{det}}[\phi]-\int\!
  d^{3}zd^3z'
  J_{i}^{\dagger}(z)G^{i}_{~j}[\phi](z;z')J^{j}(z')\right),
\end{align}
with 
\begin{align}
  S_b &= \int\!\mathrm{d}x\mathrm{d}y \mathrm{d}\tau \left[
    \frac{1}{2}\left(\partial_{\tau}\phi\right)^{2}+\left(\partial_x\phi\right)^{2}+\frac{1}{2}\left(\partial_y\phi\right)^{2}
  \right] \non
S_{\text{det}} &= \int\! \mathrm{d}x\mathrm{d}y\mathrm{d}\tau \left[- N_f\text{Tr}\ln
  G{}^{-1}[\phi] \right]
\label{eq:Sdet}
\end{align}
and $G^{i}_{~j}[\phi](z;z')=\delta^{i}_{j}G[\phi](\tau,x,y;\tau',x',y')$ is the fermionic propagator
in presence of a background bosonic field configuration. By definition
it satisfies
\begin{align}
\label{eq:2}
  \left(-\partial_{\tau} +iv\partial_x+\lambda \phi(\tau,x,y)\right) G[\phi](\tau,x,y;\tau',x',y') =\delta(\tau-\tau')\delta(x-x')\delta(y-y') 
\end{align}
Taking functional derivatives with respect to the sources, the full
fermion Green's function is then given by a path integral over only
the bosonic field:
\begin{equation}
\langle \psi^{\dagger}_j(z)\psi^i(0)\rangle_{\text{exact}} =
\delta^{i}_jG(z,z')=\delta^{i}_j \frac{\int\!\cD\phi\,G[\phi](z,z')e^{-S_{b}[\phi]-S_{\text{det}}[\phi]}}{\int\!\cD\phi\,e^{-S_{b}[\phi]-S_{\text{det}}[\phi]}}.\label{eq:pathint}
\end{equation}

\subsection{The $N_f=0$ quenched approximation and Landau damping}

We will evaluate this integral in the
quenched 
or Bloch-Nordsieck
approximation
. This is a well known ad hoc approximation in
lattice gauge theory  \cite{Golterman:1994mk}
and finite temperature QED \cite{BN1, BN3, BN4, Karanikas:1992aj}
whereby all contributions from $S_{\text{det}}$ are ignored: one
  sets the one-loop (fermion) determinant to one by hand. In our context we can
  make this approximation precise. Eq.~\eqref{eq:Sdet} shows that
  $S_{\text{det}}$ is directly proportional to $N_f$, whereas no other
  terms are. From Eq.~\eqref{eq:pathint} it is then clear that this approximation computes the
  leading contribution to the full fermion Green's function in the
  limit $N_f\rar 0$. Note that we consider the $N_f$ limit within
    correlation functions and not directly in the partition function.

Diagrammatically this means that one 
  considers only contributions to the full Green's functions that do
  not contain fermion loops. Fermion loop corrections to the bosonic
  propagator, however, encode the physics of Landau
  damping. 
As discussed, this is important in the deep infrared and requires treatment of the dangerously irrelevant quadratic corrections to the kinetic term
$\mathcal{L}_{\text{curv}}$ due to Fermi surface curvature. It is its
  Landau
  damping contribution that redirects the RG flow. 
As we explained in the introduction there is believed to be a significant 
 intermediate energy regime where the damping can be neglected \cite{Stanford4, Allais:2014fqa}. 
This is the regime
captured by the $N_f \rar 0$ limit. Since this limit tames the
  dangerous nature of irrelevant Fermi surface curvature,  this also justifies our patch
approximation and linearization of the fermion dispersion relation. We focus on this
regime in this paper. The effects of Landau damping are precisely captured by
the $\cO(N_f)$ corrections; we leave these for future research.

We will now show how
 in this intermediate regime captured by the 
quenched approximation, where one may freely ignore Fermi surface
  curvature and use a planar ``patch'' dispersion,
$G_{\text{full}}(z;z')$ can be determined exactly.
This is because the fermion two-point function in the presence of a
background field $G[\phi](z;z')$ depends on the background bosonic
field exponentially. The overall path integral over $\phi$ therefore
remains Gaussian even in presence of the Yukawa interaction.

\subsection{The exact fermion Green's function}

First, we determine the fermion Green's
function in the presence of an external boson field $G[\phi]$. Rather
then working in momentum space, it will be much more convenient to
work in position space. 
Note that because the background scalar field $\phi(\tau,x,y)$ can be
arbitrary, the fermionic Green's function $G\left[\phi\right]$ is not
 translationally symmetric. However, translational invariance will be
 restored after evaluating
 the path integral over $\phi$.

Rewriting the background dependent Green's function as 
\begin{align}
G[\phi]&\left(\tau_{1},x_1,y_1;\tau_{2},x_2,y_2\right)=\non
&\tilde{G}_{0}\left(\tau_{1}-\tau_{2},x_{1}-x_{2},y_1-y_2\right)\exp\left(-\lambda V[\phi]\left(\tau_{1},x_1,y_1;\tau_2,x_{2},y_{2}\right)\right),\label{eq:solution}
\end{align}
with $\tilde{G}_{0}$ the translationally invariant free Green's function in real space
\begin{align}
\tilde{G}_{0}(\tau,x,y)&=-\frac{i}{2\pi}\frac{\mbox{sgn}(v)}{x+iv\tau}\delta(y)
\equiv G_0(\tau,x)\delta(y)
,\label{eq:free}
\end{align}
it is readily seen that the solution to the defining Eq.~\eqref{eq:2}
is given by
\begin{align}
&V[\phi]\left(\tau_{1},x_1,y_1;\tau_2,x_{2},y_{2}\right)=\non
&\int\! \mathrm{d}x\mathrm{d}y\mathrm{d}\tau \left[\tilde{G}_{0}\left(\tau_1-\tau,x_{1}-x,y_{1}-y\right)-\tilde{G}_{0}\left(\tau_2-\tau,x_{2}-x,y_{2}-y\right)\right]\phi(\tau,x,y), \label{eq:exponent}
\end{align}
To be more precise, we need to ensure that the background dependent Green's function (\ref{eq:solution})
satisfies proper boundary conditions as well. We did so by considering the problem at finite temperature and volume
and taking explicitly the continuum limit. The compact analog of (\ref{eq:solution})
has to satisfy antiperiodic boundary condition along the imaginary
time direction and periodic boundary condition along
the spatial direction. This can be achieved by taking the periodic
free fermion Green's function ($\tilde{G}_0^{P}$) in the exponent
Eq. \ref{eq:exponent} and the antiperiodic one ($\tilde{G}_0^{AP}$) in
Eq. \ref{eq:solution}. In the continuum limit, however, their
  functional forms are indistinguishable, and we denote them
  with the same symbol ($\tilde{G}_0$).

The insight is that the only dependence on $\phi$ in the background
dependent Green's function is in the exponential factor $V[\phi]$ and
that this dependence is linear. In combination with the quenched $N_f
\rar 0$ limit, the path-integral over $\phi$ Eq. (\ref{eq:pathint}) 
needed to obtain the full
Green's function is therefore Gaussian, and we can 
straightforwardly evaluate this to (\ref{eq:pathint}) to obtain
\begin{equation}
\label{eq:3}
G\left(\tau_1,x_1,y_1;0\right)=G_{0}\left(\tau_1,x_1\right)
  \delta(y_1)\exp\left[ I(\tau_1,x_1;0)\right]
\end{equation}
with
\begin{align}
&I(\tau_1,x_1;0)=\non
&\frac{\lambda^{2}}{2}\int\!\! \text{d}x\text{d}\tau \text{d}x' \text{d}\tau' 
M(\tau_1-\tau,x_1-x;-\tau,-x) G_B(\tau-\tau',x-x',0)
                 M(\tau_1-\tau',x_1-x';-\tau,-x),
\end{align}
where
\begin{align}
M(\tau_1,x_1;\tau_2,x_2) &=
                           G_0(\tau_1,x_1)-G_0(\tau_2,x_2)~, \label{eq:exactres}
\end{align}
and $G_B(\tau-\tau',x-x',y-y')=G_B(\tau,x,y;\tau',x',y')$ equal to the translationally invariant free boson
propagator defined by
\begin{align}
  \left(\pa_\tau^2+\nabla^2\right)G_B(\tau,x,y;\tau',x',y') = -\delta(\tau-\tau')\delta(x-x')\delta(y-y')~.
\end{align}

Eq.~\eqref{eq:3} is a remarkable result. In the $N_f\rar 0$ quenched
approximation the full fermion Green's function still consists
of a complicated set of Feynman diagrams that are normally not
resummable. In particular at the two-loop level there are rainbow diagrams
(Fig.~\ref{fig:two-loop}(c)) and
vertex-corrections of self-energies (Fig.~\ref{fig:two-loop}(b))
that do not readily combine to a summable series. The
reason why in this $N_f\rar 0$ planar patch theory we can do so, is
the existence of the following multiplicative identity of fermion
propagators in the planar limit where the dynamics is effectively 1+1
dimensional.
\begin{align}
  G_0(\tau_1,x_1)G_0(\tau_2,x_2)=G_0(\tau_1+\tau_2,x_1+x_2)\left(G_0(\tau_1,x_1)
  + G_0(\tau_2,x_2) \right)
\end{align}
This identity follows directly from trivial equality
\begin{align}
   (G_0(\tau_1,x_1))^{-1}+(G_0(\tau_2,x_2))^{-1}=(G_0(\tau_1+\tau_2,x_1+x_2))^{-1},
\end{align}
and 
has many corollary multiplicative identities for
products of $n>2$ planar fermion propagators.
The usual perturbative series and the exact
result Eq.~(\ref{eq:exactres}) may seem different but their equality can be proven to all orders. We do so in
Appendix \ref{sec:append-a:-comp}, thereby unambiguously establishing
that this is the {\em exact} fermion two-point function in the planar
theory in the
quenched approximation.

\section{The physics of the planar quenched quantum critical metal}
\label{sec:results} 

We now show that this all order result for the fermion Green's
function, albeit in the quenched $N_f\rar 0$ approximation, describes
very special physics. In this approximation the fermionic excitations
constitute a continuous spectrum of excitations with power-law tails
analogous to a critical theory. Importantly, in the low energy limit
this continuous spectrum centers at three distinct momenta
with different exponents for the
power-law fall-off.

To exhibit this exotic physics from the exact $N_f\rar 0$ Green's
function \eqref{eq:3}, we substitute the explicit form of the
  boson and fermion Green's functions and Fourier transform
the internal integrals. For the exponent $I(\tau,x;0)$ we then  have:
\begin{equation}
I(\tau,x;0)=\frac{\lambda^2}{8\pi^3}\int \! \mathrm{d}\omega\mathrm{d}k_x\mathrm{d}k_y\frac{\cos(\omega \tau-k_x x)-1}{(i \omega-v k_x)^2(\omega^2+k_x^2+k_y^2)}
\label{integral1}
\end{equation}
This integral can be done analytically to obtain (for $v^2\neq1$)
\begin{equation}
I(\tau,x;0)=\frac{\lambda^2}{8\pi(1-v^2)}\left(\frac{(\tau -i v x)}{\sqrt{1-v^2}} \log \left(\frac{\tau
   -i v x+\sqrt{\left(1-v^2\right) \left(\tau ^2+x^2\right)}}{\tau -i v x-\sqrt{\left(1-v^2\right) \left(\tau
   ^2+x^2\right)}}\right)-2 \sqrt{\tau ^2+x^2}\right);
\end{equation}
for $v^2=1$ one obtains
\begin{equation}
I_{v^2=1}(\tau,x;0)=\lambda^2\frac{(\tau+i\sgn(v)x)^2}{12\pi \sqrt{\tau ^2+x^2}}.
\end{equation}
This gives us the all order $N_f\rar 0$ Green's function in real
space. Note that this all-order Green's function
  surprisingly does not depend on any UV-regulator, despite the fact
  that it sums an infinite number of loops. The one internal integral remaining in
  Eq.~(\ref{integral1}) is finite.

\bigskip

Analytically continuing in $\tau$ for $0<v<1$ yields the retarded
Green's function.
The physics follows from Fourier transforming this real time Green's
function to momentum space; this is described in
Appendix \ref{appendixFourierTransform}. The resulting retarded Green's function in
momentum space is given by
 \begin{equation}
\begin{split}
G_R(\omega,k_x)=\frac{1}{\omega -k_x v+\frac{\lambda ^2 }{4\pi\sqrt{1-v^2}}\sigma(\omega,k_x)}, \label{eq:retarded}
\end{split}
\end{equation}
where $\sigma(\omega,k_x)$ is the root, within $0<\im(\sigma)<i\pi$, of the equation
\begin{equation}
\frac{\lambda ^2 }{4\pi\sqrt{1-v^2}}(\sinh (\sigma)-\sigma\cosh (\sigma))+ v \omega-k_x-\cosh (\sigma)
   (\omega-k_x v +i \epsilon )=0
   \label{implicit1}
\end{equation}
A small positive parameter
$\epsilon$ is introduced to identify the correct root when it otherwise would be on the real axis, which is the case for
\begin{equation}
(k_x+\omega ) \left(\lambda ^2 \sinh \left(\frac{4 \pi  \sqrt{1-v^2} (v k_x-\omega
   )}{\lambda ^2}\right)+4 \pi  \sqrt{1-v^2} (v \omega -k_x)\right)\geq 0.\label{real}
\end{equation} 
Note that the center-combination $v\omega-k_x$ is correct; we are
working in units in which the boson-dispersion velocity $c=1$. The units
can be made correct by restoring $c$.

\bigskip
Expression \eqref{eq:retarded} together with \eqref{implicit1} is the main technical result of the paper.
We can now extract the insights into the spectrum of 
fermionic excitations around the ground state of the planar quenched
metal. Fig. \ref{spectral} plots the spectral function
$A(\omega,k_x)=-2\mbox{Im} G_R(\omega,k_x)$ as a function of the
dimensionless combinations $\omega/\lambda^2,~k_x/\lambda^2$, as
$\lambda$ is the only scale in the problem. We immediately note that
there is an obvious continuous peak, corresponding to a clear
excitation in the spectrum. This excitation has the properties that: 

\begin{figure}[t]
\includegraphics{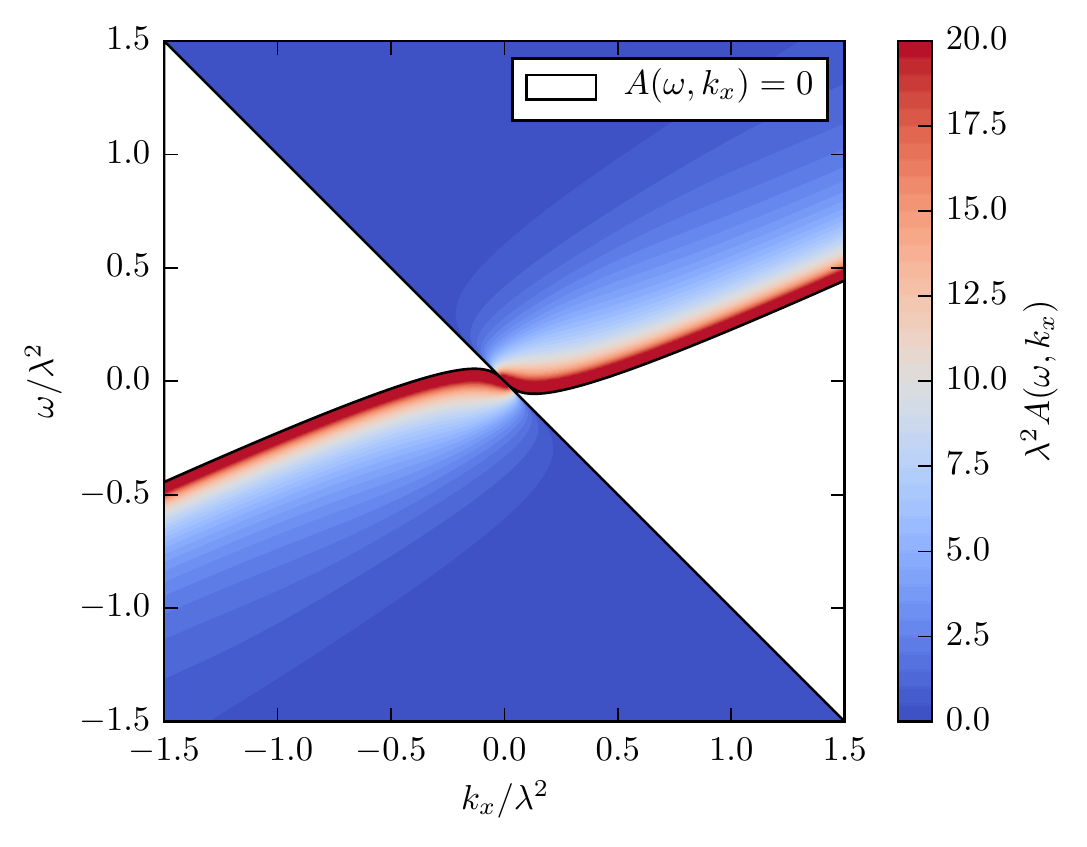}
\protect\caption{Fermionic spectral function for $v=0.5$. It is identically zero in the white region.}
\label{spectral}
\end{figure}

\begin{figure}[t]
\includegraphics{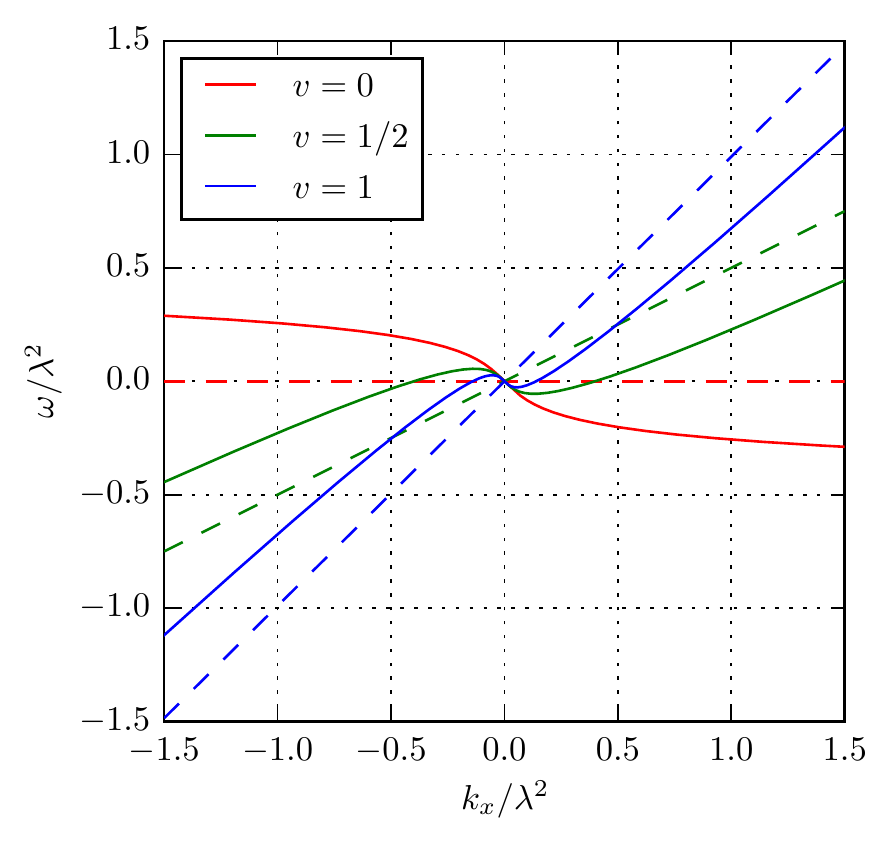}\protect\caption{The dispersion relation (the zeros of $G^{-1}$) for different $v$ in
the interacting theory (solid line) and the free theory (dashed)}
\label{dispersion}
\end{figure}

\begin{itemize}
\item The dispersion relation is $S$-shaped in the infrared near
  $k_x=0$, and now has {\em three} intersections with
  $\omega=0$. A truncation of the theory to very low energies would
  therefore indicate three distinct Fermi surfaces. Similar topological
Fermi surface instabilities due to electron interaction have been found
e.g. in \cite{TopInst}. Curiously the dispersion is nearly identical to the
  one-loop result.

\item As has been demonstrated before by means of a perturbative
  renormalization group analysis \cite{Stanford2}, we see the speed of fermions $v$
decreases as we go from high to low frequency/momentum. The distinct
$S$-shaped curve is outside of the regime of perturbation theory,
however. With the exact result we see that the emergent Fermi-velocity at the innermost /new{($k_x=-k_x^{\ast}<0$) and the outermost ($k_x=k_x^{\ast}>0$)}
Fermi surfaces is non-universal, but positive and depends on the UV
fermionic velocity $v$. These Fermi-surfaces are therefore
particle-like.

However, the reverse of direction due to the $S$-shape shows that the
Fermi velocity at the
emergent Fermi-surface at the original Fermi-momentum $k_x=0$ is now
in the opposite direction and the surface is therefore 
hole-like. Moreover, the value of the emergent Fermi-velocity at
$k_x=0$ is universal: it equals the boson-velocity $v_{F}=-1$ at $k=0$ (near the middle Fermi
surface), independent of the UV
fermionic velocity $v$ (Fig. \ref{dispersion}). A way to perceive what
happens is that the hole-like
excitations at $k_x=0$ become tied to the critical boson which
completely dominates the dynamics.

\item The new Fermi surfaces are symmetric around $k_x=0$; the
    hole-like one is at $k_x=0$ and
as follows from Eq. \eqref{eq:retarded} and \eqref{implicit1} the two
particle-like ones are symmetrically arranged at  $\pm k_x^{\ast}$. The
precise value of $k_x^{\ast}$ depends on
the initial fermi velocity $v$. In the planar approximation where the
Fermi surface is infinite in extent, this guarantees that Luttinger's
theorem holds: the original Fermi surface (the region $-\infty < k_x <0$)
has the same volume as the emergent two regions enclosed by Fermi surfaces ($-\infty < k_x <-k_x^{\ast}$ and $0<k_x <k_x^{\ast}$).

\item The spectral function, $A(\omega,k_x)=-2\mbox{Im} G_R(\omega,k_x)$,
is identically zero for the range of $\omega$ and $k_x$ whenever
$\sigma(\omega,k_x)$ is exactly real. This is whenever
Inequality \eqref{real} is satisfied. Such a large range of zero-weight may
seem to violate unitarity. As a consistency check, however, it can be
demonstrated that the
 Green's function satisfies the sum rule for all $v$ (Appendix \ref{sec:integr-spectr-funct})
 \begin{equation}
\int\limits_{-\infty}^\infty \mathrm{d}\omega A(\omega,k_x)=2\pi,\,\,\forall k_x  
 \end{equation}

\item 
Importantly, the weight of
  the spectral function is infinite {\em at all points of
    the dispersion relation}. Substituting the implicit dispersion relation
  $\sigma = \frac{4\pi\sqrt{1-v^2}}{\lambda^2}(\omega-vk_x)$ into the constraint
    Eq.~\eqref{implicit1}, one can verify this explicitly.
The spectrum is therefore a continuum, and not discrete. The
excitation spectrum therefore resembles that of a scale-invariant
critical theory, rather than that of interacting particles.


\item
Focusing on the low-energy regime, i.e. a narrow band in the spectral
function around $\omega =0$, we can determine the spectral weight analytically around the three different Fermi-surfaces
--- the three different crossings of the dispersion relation with
$\omega=0$. Expanding Eq. \eqref{implicit1} around $(\omega,k_x)=(0,0)$ the retarded Greens's function behaves as 
\begin{equation}
G_{R}\left(\omega,k_{x}\right)=C\lambda^{-4/3}\left|\omega+k_{x}\right|^{-1/3},
\end{equation}
with
$C=\left(1+v\right)^{1/3}\frac{4\pi}{(12\pi)^{1/3}}\left[i\frac{\sqrt{3}}{2}-\frac{1}{2}\sgn\left(\omega+k\right)\right]$, 
whereas near the outer Fermi surfaces $(\omega,k_{x})=(0,\pm k_{x}^{*})$ we have
\begin{equation}
G_{R}\left(\omega,k_{x}\right)\sim\lambda^{-1}\left(v^*k_{x}-\omega\right)^{-1/2}.
\end{equation}
In each case the IR $\omega \simeq 0$ spectral function thus has a clear power-law
behavior with a branch-cut singularity, but it has a different
exponent depending on the Fermi surface. Furthermore, at the $k_x=0$
Fermi surface the spectral function
is symmetric around $\omega$, while in the other two it is zero
for negative (positive) frequencies. This is clearly visible in
Fig. \ref{spectral}.

Interestingly, in all three cases the power-law scaling conforms with a
uniform scaling of energy and  momentum corresponding to groundstate
with a dynamical
critical exponent $z_f=1$ (consistent with \cite{Stanford1, Stanford4}). This is in contrast to the expectation
that the 2+1 dimensional quantum critical metal has a $z_f \neq 1$
groundstate \cite{Sachdev1}. However, the role of Landau damping and
Fermi surface curvature is crucial in this expectation, and both are
ignored in the planar $N_f=0$ approximation here.
\end{itemize}

\begin{figure}
\includegraphics{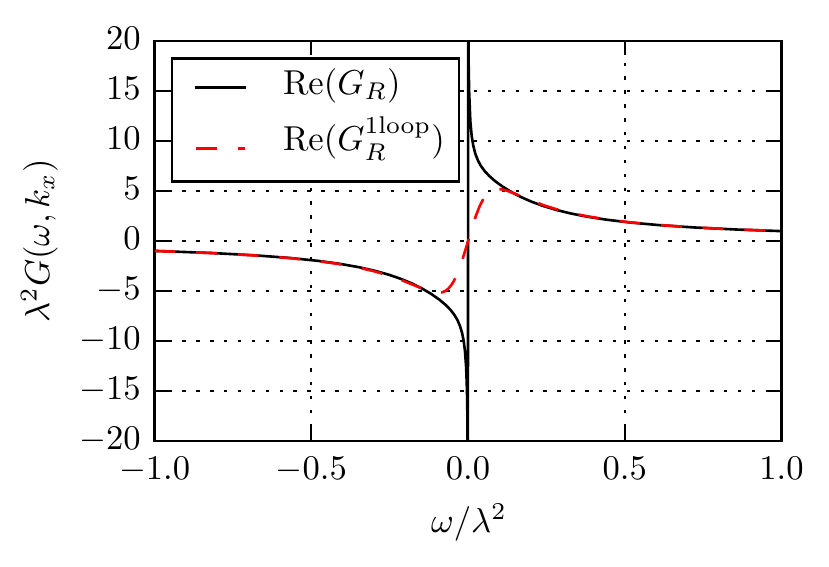}\includegraphics{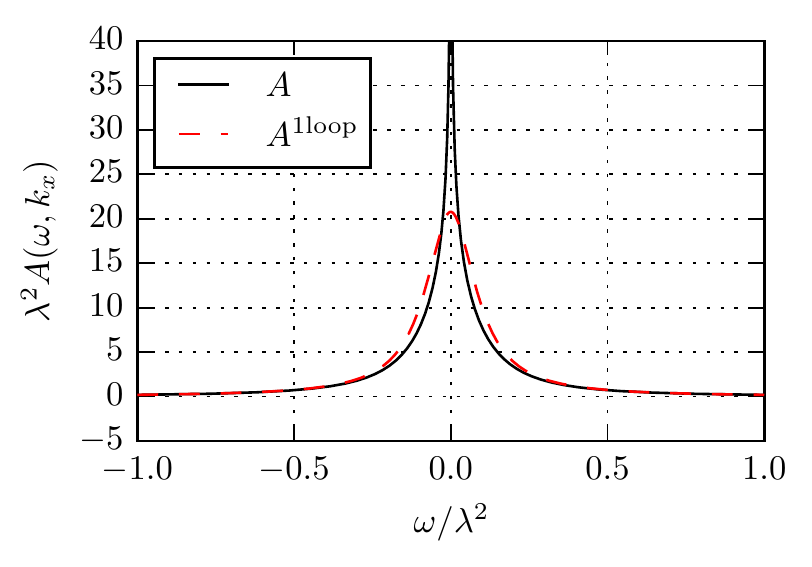}
\protect\caption{The real and the imaginary part of the Green's function at the middle Fermi surface ($k_{x}=0$, $v=0.5$) as a function of $\omega$
(solid line), compared with the corresponding one-loop result (dashed)}
\label{inner}
\end{figure}

\begin{figure}
\includegraphics{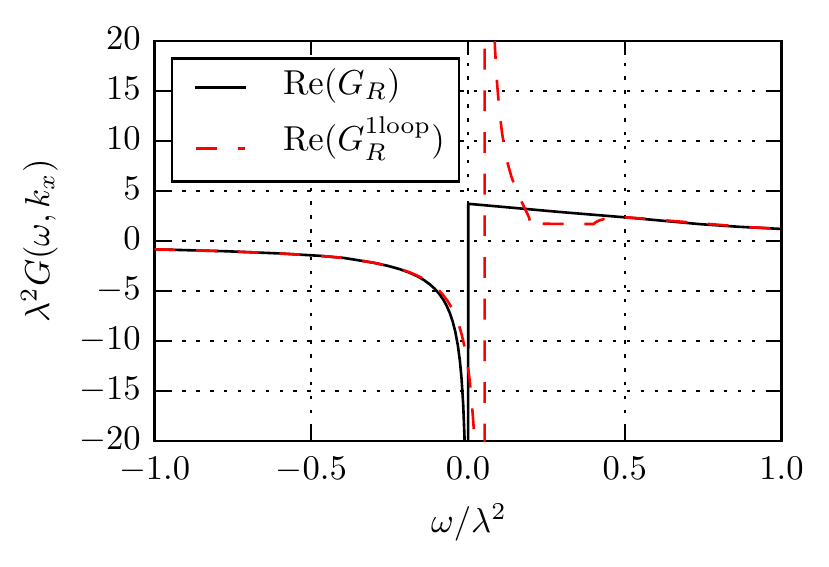}\includegraphics{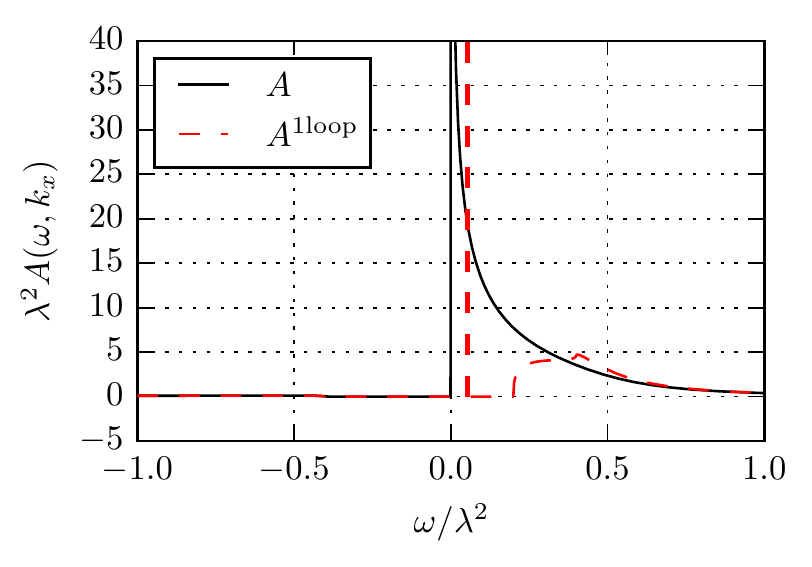}
\protect\caption{The real and the imaginary part Green's function at the outer Fermi surface ($k_{x}=k_x^*\approx 0.4\lambda^2$, $v=0.5$)
as a function of $\omega$
(solid line), and the one-loop truncated result (dashed). The spectral weight of the one-loop approximation is concentrated to a $\delta$-function whereas the spectral weight of the full result is spread in the power-law singularity.}
\label{outer}
\end{figure}

All these insights are non-perturbative. This can be readily shown by
comparing our exact result to the one-loop perturbative
answer (Fig. \ref{inner} and Fig. \ref{outer}). The
one-loop result is only a good approximation in the UV, far away from the continuous set of excitations, i.e. the dimensionful
Yukawa coupling
$\lambda^2 \ll \lvert \omega - v k_x\lvert$. Perturbation theory therefore fails to capture any of the
distinct non-Fermi liquid phenomenology of the IR (with the exception of the shape of the dispersion-curve).

For completeness we can also compute the density of states and the occupation number as a function of momentum. The former gives (see Appendix \ref{sec:integr-spectr-funct})
 \begin{align}
 N(\omega)= \int\!\mathrm{d}k_x \frac{A(\omega,k_x)}{2\pi} = \frac{1}{v}\left[1+\theta(\lambda^2\frac{\cosh ^{-1}\left(v^{-1}\right)-\sqrt{1-v^2}}{4\pi(1-v^2)^{3/2}}-\lvert\omega\rvert)\right].
 \end{align}
 
\begin{figure}
\includegraphics{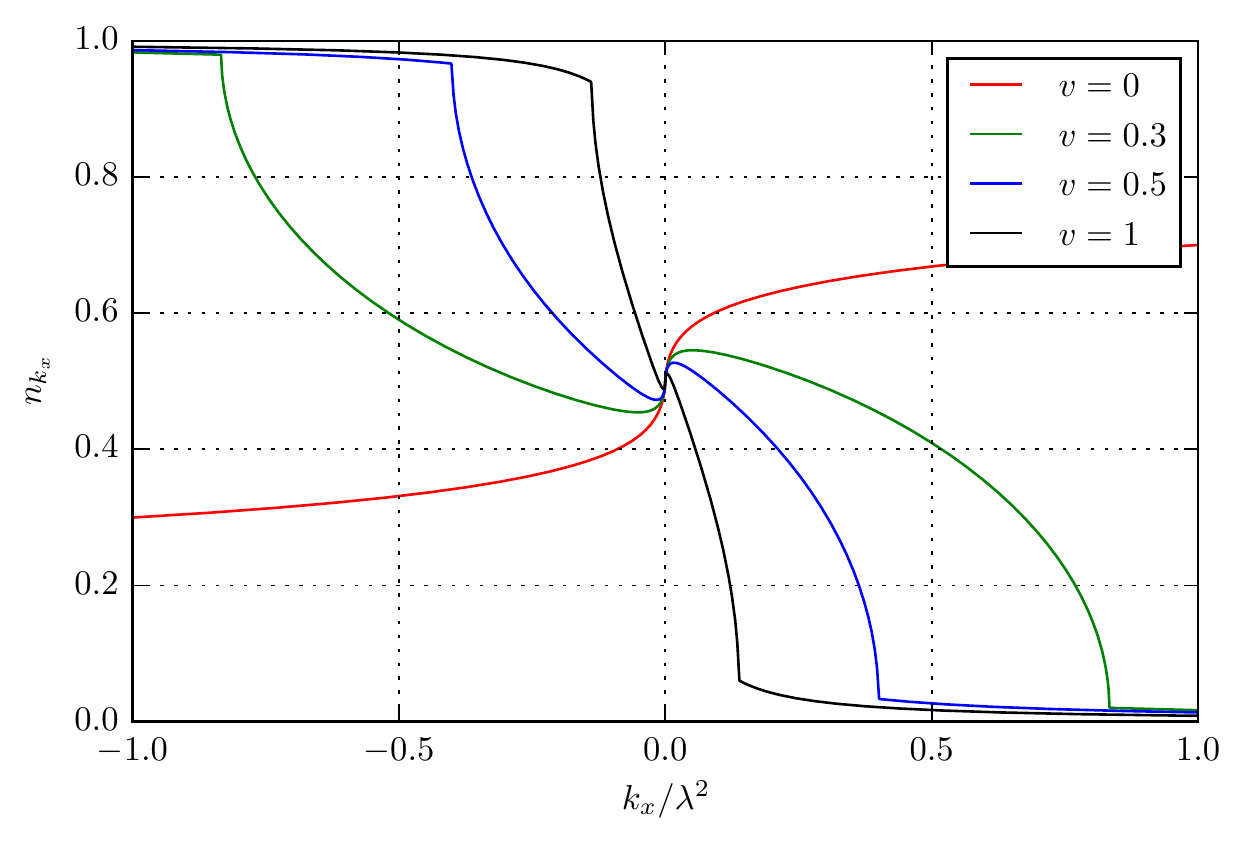}

\protect\caption{The occupation number for different $v$ as a function of momentum.
We can see the effect of the multiple sharp fermi surfaces, the non-fermi
liquid behaviour is also apparent as a non-monotonic behaviour of the occupation number.}
\label{occupation}

\end{figure}

The result for the occupation number is (Appendix \ref{sec:integr-spectr-funct})
\begin{align}
n_{k_{x}}=\int_{-\infty}^{0}\!\!\mathrm{d}\omega\frac{A(\omega,k_{x})}{2\pi}=
 \frac{1}{\pi}\arg\left[
 \frac{  \cosh(\sigma(\omega=0, k_x/\lambda^2, v))-v }{ v  \cosh^{-1}(v)   -\sqrt{1 - v^2} (i - 4 \pi (1-v^2)k_x/\lambda^2)  }\right],
  \end{align}
where $\sigma(\omega,k_x)$ is defined by \eqref{implicit1}. It is plotted in Fig. \ref{occupation}. 
We can see the effect
of the multiple Fermi surfaces as discontinuities in the derivative of
the occupation number, even though the occupation number itself is
continuous. This is another way to see that the fermionic
excitation spectrum is that of a non-Fermi liquid. (Note that in the
singular case of vanishing UV Fermi velocity $v=0$, the occupation
number has different asymptotics as $k_x\rightarrow\pm\infty$ than for
any small but non-zero $v$. For $v\neq 0$ the occupation number approaches
 $n_{k_{x}}= 1$ at $k_{x}=-\infty$ and $n_{k_{x}} =0$ at $k_x=\infty$, as it should.)


\section{Conclusions} 
\label{sec:conclusions}

In this paper we have shown that in the quenched $N_f\rightarrow 0$ limit the fermion Green's function in a 2+1 dimensional quantum critical metal can be determined exactly. The quenched limit neuters the dangerous nature of dimensionally irrelevant Fermi surface corrections and allows us to truncate to a linear dispersion relation for the fermions. This reduction to an effective one-dimensional system allows an explicit solution to the fermion Green's
function in the presence of a background scalar field. The quenched $N_f \rightarrow 0$ limit further allows us to compute the full background scalar field path-integral when coupled minimally to the fermion.

Even though the quenched limit discards the physics of Landau damping, our result shows that the resulting physics is already very non-trivial. There are three distinct low-energy excitations as opposed to the excitations around a single Fermi surface of the free theory. Most importantly, the sharp excitations of the free theory broaden into a power-law singularity of the spectral function of the form $G\sim(\omega-\epsilon(k_x))^{-\eta}$, with either $\eta=1/2$ or $\eta=1/3$. The groundstate is a non-Fermi-liquid.

Beyond the quenched limit and including $N_f$ corrections, i.e. fermion loops, Landau damping effects become important. 
These effects will show up below some energy scale $E_\mathrm{LD}$ set by both $N_f$ and the Fermi surface curvature $\kappa$. Our model breaks down below this scale, 
but it is expected to describe the physics above $E_\mathrm{LD}$. Qualitatively the physics is that of a non-Fermi-liquid both above $E_\mathrm{LD}$ and below $E_\mathrm{LD}$ \cite{Altshuler2}, but in detail it will differ.

In order to access IR physics below $E_\mathrm{LD}$, the corrections in the Fermi surface curvature and the number of fermionic flavours must be treated systematically, but
a (possible) shortcut deserves to be mentioned. Our analytic determination of the exact fermionic Green's analytically hinged on the free fermion 
dispersion being linear, but the approach taken in this paper does not put any restrictions on the allowed form of the bosonic propagator. This opens up the possibility to implement the Landau-damping effects phenomenologically,
just by modifying the background bosonic Green's function, and staying within the Gaussian approximation. This is the approach taken by Khveshchenko and Stamp \cite{Khveshchenko1} and Altshuler, Lidsky, Ioffe and Millis \cite{Altshuler1,Altshuler2}. Comparing to vector large $N_f$ approaches \cite{Metlitski1,Metlitski2}, it is not clear that this is sufficient to reliably capture the IR. The Landau damping is not the only important effect. Interactions of the boson field with itself beyond the Gaussian approximation must also be taken into account, e.g. our model needs to be enhanced by a $\phi^4$ interaction to describe the Ising-nematic critical point \cite{Allais:2014fqa}. We leave study of the related effects
for a subsequent paper.

The interesting question will be which non-Fermi liquid features are retained and which change. The dynamical critical exponent $z_f$ below the Landau damping scale is likely different from 1. Also, the splitting of the Fermi surface seems to be a subtle phenomenon, and whether it remains stable upon including fermionic loop corrections or going beyond the local patch approximation requires a careful investigation. On the other hand, the destruction of the quasi-particle poles and the fact that the spectrum is singular along the full dispersion curve is expected to be a robust effect that resembles that of a critical state. This is thought to be enhanced by the Landau damping.



\acknowledgements
We are greatful to Mikhail Titov, Mikhail Katsnelson and Jan Zaanen for discussions. This work was supported in part
by a VICI (KS) award of the
Netherlands Organization for Scientific Research (NWO), by the
Netherlands Organization for Scientific Research/Ministry of Science
and Education (NWO/OCW), by a Huygens Fellowship (BM), and by
the Foundation for Research into Fundamental Matter (FOM).
KS expresses his sincerest thanks to the Harvard department of
Physics, where this project was started.

\appendix

\section{Comparison with perturbation theory}
\label{sec:append-a:-comp}

We can expand (\ref{eq:exactres}) in the coupling constant. Although
at first sight this expansion seems different from the usual perturbative
expansion, we will show that in the case of zero fermi surface curvature
they match at any order if we do not include fermion loops.

The $\lambda^{2n}$ term in (\ref{eq:exactres}) is
\begin{equation}
G_{n}(z)=\frac{G_{0}(z)}{2^{n}n!}\left(\int dx'dx''d\tau'd\tau''\left[G_{0}\left(z-z'\right)+G_{0}\left(z'\right)\right]G_{B}\left(z'-z''\right)\left[G_{0}\left(z-z''\right)+G_{0}\left(z''\right)\right]\right)^{n},
\end{equation}
where $z=x+iv\tau$. The usual perturbative expansion result can be
obtained by expanding
\begin{equation}
\langle\psi(z)\psi(0)^{+}\exp(\lambda\phi\psi^{+}\psi)\rangle
\end{equation}
and evaluating by Wick contraction
\begin{equation}
G^{\mathrm{pert}}_{n}(z)=\frac{(2n-1)!!}{(2n)!}\int
dx_{1}...dx_{2n}d\tau_{1}...d\tau_{2n}I\cdot
G_{B}(x_{1}-x_{2},\tau_{1}-\tau_{2})...G_{B}(x_{2n-1}-x_{2n},\tau_{2n-1}-\tau_{2n}),
\nonumber
\end{equation}
\begin{equation}
I=\sum_{(i_{1},..,i_{2n})\in S_{2n}}G_{0}\left(z-z_{i_{1}}\right)G_{0}\left(z_{i_{1}}-z_{i_{2}}\right)...G_{0}\left(z_{i_{2n-1}}-z_{i_{2n}}\right)G_{0}\left(z_{i_{2n}}\right)
\end{equation}

Here $S_n$ is the set of permutations of the numbers 1 through $n$. The factor $1/(2n!)$ comes from the Taylor expansion of the exponential.
By summing over the different assignments of internal points we are explicitly
counting the different contractions of the fermion fields.
There are however still $(2n-1)!!$ possibilities to pair the boson
fields (each pairing gives rise to the same contribution after a change
of variable in the integral). Since $(2n)!/(2n-1)!!=n!2^{n}$ the
identity which remains to be proved, once we have used our simple form of the free fermion Green's function, is
\begin{equation}
\sum_{(i_{1},..,i_{m})\in S_m}\frac{1}{z-z_{i_{1}}}\frac{1}{z_{i_{1}}-z_{i_{2}}}...\frac{1}{z_{i_{m-1}}-z_{i_{m}}}\frac{1}{z_{i_{m}}}=\frac{z^{m-1}}{\left(z-z_{1}\right)\left(z-z_{2}\right)...\left(z-z_{m}\right)z_{1}...z_{m}}.\label{eq:tobeproved}
\end{equation}
We need this for $m=2n$, but the statement is true for odd $m$
as well.

The identity can be proven by induction. The $m=1$ case is easily checked and given that the equality holds for $m-1$ we have
\begin{equation}
\sum_{(i_{1},..,i_{m})\in S_m}\frac{1}{z-z_{i_{1}}}\frac{1}{z_{i_{1}}-z_{i_{2}}}...\frac{1}{z_{i_{m-1}}-z_{i_{m}}}\frac{1}{z_{i_{m}}}=\frac{1}{z_{1}...z_{m}}\sum_{k=1}^m\frac{1}{z-z_{k}}\frac{z_{k}^{m-1}}{\left(z_{k}-z_{1}\right)\left(z_{k}-z_{2}\right)...\left(z_{k}-z_{m}\right)}.\label{eq:simplification}
\end{equation}
where the product in the last denominator excludes $(z_k-z_k)$.
The right hand side of (\ref{eq:tobeproved}) and (\ref{eq:simplification})
are the same since they are both meromorphic functions of $z$ with the
same pole locations and residues and they both approach 0 at $\infty$.

\section{Calculating Real-Space Fermion Green's
  Function}
\label{appendixIntegralInExponent}

To find the real-space Euclidean fermionic Green's function we have to evaluate the integral \eqref{integral1}. 
In order to do that, it is convenient to firstly make a coordinate transformation of the following form 
\begin{equation}
\begin{split}
\omega=\frac{x k_1+\tau k_2}{\sqrt{x^2+\tau^2}},\\
k_x=\frac{x k_1-\tau k_2}{\sqrt{x^2+\tau^2}}.
\end{split}
\end{equation}
The integral in $k_2$ can be then explicitly evaluated, giving
\begin{equation}
I=\int \mathrm{d}k_1\mathrm{d}k_y\frac{\lambda^2  \left(\tau ^2+x^2\right) \left(\cos \left(k_1 \sqrt{\tau
   ^2+x^2}\right)-1\right)}{8\pi^2\sqrt{k_1^2+k_y^2} \left(x
   \left(\sqrt{k_1^2+k_y^2}+\lvert k_1\rvert v\right)-i \tau  \left(v
   \sqrt{k_1^2+k_y^2}+\lvert k_1\rvert\right)\right)^2}.
   \end{equation}
   Now switching to polar coordinates, $k_1=k\cos\theta,\,k_y=k\sin\theta$, and performing the radial integral in $k$ we obtain
\begin{equation}   
  I=\int_0^{2\pi} \mathrm{d}\theta \frac{\lambda^2 \lvert\sin (\theta )\rvert \left(\tau ^2+x^2\right)^{3/2}}{16\pi (\tau  v+ix+\lvert\sin(\theta )\rvert (i v x+\tau
   ))^2}.
\end{equation}   
Finally, integrating over $\theta$ for $v^2\neq1$ we derive
\begin{equation}
I=\frac{\lambda^2}{8\pi(1-v^2)}\left(\frac{(\tau +i v x)}{\sqrt{1-v^2}} \log \left(\frac{\tau
   +i v x+\sqrt{\left(1-v^2\right) \left(\tau ^2+x^2\right)}}{\tau +i v x-\sqrt{\left(1-v^2\right) \left(\tau
   ^2+x^2\right)}}\right)-2 \sqrt{\tau ^2+x^2}\right). \label{eq:v<1}
\end{equation}

For the specific case $v^2=1$ the integration should be done independently and gives a simpler result
\begin{equation}
I=\lambda^2\frac{(\tau-i\sgn(v)x)^2}{12\pi \sqrt{\tau ^2+x^2}}.
\end{equation}

To Fourier transform the corresponding Green's function to momentum space, we will need an analytical continuation.
For $0<v<1$, \eqref{eq:3} 
 with exponent \eqref{eq:v<1} can be analytically continued in $\tau$ to the complex plain 
with two branch-cuts along parts of the imaginary axis
 $\tau\in\mathbb{C}\setminus\{\re(\tau)=0,\lvert\im(\tau)\rvert>\vert x\rvert\}$ by
 \begin{equation}
 G_E(x,\tau)=-\frac{i}{2\pi}\frac{\mbox{sgn} (v)}{x+iv \tau}e^{\frac{\lambda ^2}{8 \pi  \left(1-v^2\right)} \left(\frac{(\tau +i v x)}{\sqrt{1-v^2}} \left(i \pi  \text{sgn}(x)+2 \tanh
    ^{-1}\left(\frac{\tau +i v x}{\sqrt{\left(1-v^2\right) \left(\tau
    ^2+x^2\right)}}\right)\right)-2 \sqrt{\tau ^2+x^2}\right)}
    \label{analyticContinuation}
    \end{equation}

\section{Fourier Transforming Fermion Green's function} \label{appendixFourierTransform}

The next step is to calculate the retarded fermionic Green's function in momentum space. We know that the time-ordered momentum space Green's function of the Lorentzian signature theory, $G_T(\omega)$, is related to the Green's function of the Euclidean theory, $G_E(\omega)$, by analytical continuation
\begin{equation}
G_T(\omega,k_x)=G_E(\omega(-i+\epsilon),k_x).
\end{equation}
$G_T(\omega,k_x)$ is analytic below the real line in the left half plane and above the real line in the right half plane.
$G_E(\omega,k_x)$ is the Fourier transform in a generalized sense of \eqref{analyticContinuation}. The (rather severe) divergence at infinity has to be regularized. Since the expression we found in Appendix \ref{appendixIntegralInExponent} permitted an analytic continuation to all of the first and third quadrants, we can continuously rotate the integration contour in the Fourier transform, $\tau=t(i+\delta)$, if additionally there is a regulator analytic in the first and third quadrant. We thus have
\begin{equation}
G_T(\omega,k_x)=\int \mathrm{d} t(i+\delta) \mathrm{d} x \e^{i(\omega(-i+\epsilon)(i+\delta)t-k_x x)}G_E(t(i+\delta),x).
\end{equation}
From this we see that the real-space time-ordered Green's function is given by analytically continuing the real-space Green's function of the Euclidean theory
\begin{equation}
G_T(t,x)=iG_E(t(i+\delta),x).
\end{equation}
This slightly heuristic argument of analytical continuation in real space has been verified to give the correct Green's function up to one loop perturbation theory.
The retarded Green's function is given by
\begin{equation}
G_R(\omega,k_x)=\int \mathrm{d} t \mathrm{d} x \e^{i(\omega t-k_x x)} \theta(t)(G_T(t,x)+G^*_T(-t,-x)).
\end{equation}
$G_T(t,x)$ is of the form $t^{-1}f_1(x/t)\exp(\lambda^2tf_2(x/t))$. By performing a change of variable from $x$ to $u=x/t$ we can perform the $t$ integral. For this we need a regulator $\exp(-\epsilon t)$. The integrand of the remaining $u$ integral has compact support, $u\in[-1,1]$. 
We can perform a further change of variables
\begin{equation}
\begin{split}
\sigma&=\tanh^{-1}\left(\frac{\sqrt{(1-v^2)(1-u^2)}}{1-uv}\right).
\end{split}
\end{equation}
This function maps $[-1,v)\rightarrow\RR^+$ and $(v,1]\rightarrow\RR^+$, both bijectively. The inverse thus has two branches that we need to integrate over, one for  $u<v$ and one for $u>v$, and both integrals will be over $\RR^+$. This change of variable is consistent with the principal value integral required for the singularity at $u=v$ if the $\sigma\rightarrow\infty$ limits are performed at the same time. The integrand obtained with this change of variable can be written as a sum of four pieces
\begin{equation}
G_R(\omega,k_x)=\int_0^\infty \mathrm{d} \sigma \big [F(\sigma)+F(-\sigma)-F(\sigma+i\pi)-F(-\sigma+i\pi)\big ],
\end{equation}
where $F(\sigma)$ is defined as
\begin{equation}
F(\sigma)=\frac{i}{2\pi}\frac{\sinh (\sigma)}{\frac{\lambda ^2 (\sinh (\sigma)-\sigma\cosh (\sigma))}{4\pi\sqrt{1-v^2}}+ v \omega-k_x-\cosh (\sigma)
   (\omega-k_x v +i \epsilon )}.
\end{equation}
Since $F(\sigma)$ is a meromorphic function and it approaches 0 as $\re(\sigma)\rightarrow\pm\infty$, we can close the contour at $\pm\infty$ and obtain the integral as the residue of $F(\sigma)$'s single pole in the strip $0<\im(\sigma)<i\pi$,
 \begin{equation}
\begin{split}
G_R(\omega,k_x)=\frac{1}{\omega -k_x v+\frac{\lambda ^2 }{4\pi\sqrt{1-v^2}}\sigma(\omega,k_x)}
\label{GreenImpl}
\end{split}
\end{equation}
where $\sigma(\omega,k_x)$ is the root, within $0<\im(\sigma)<i\pi$, of the equation
\begin{equation}
0=\frac{\lambda ^2 }{4\pi\sqrt{1-v^2}}(\sinh (\sigma)-\sigma\cosh (\sigma))+ v \omega-k_x-\cosh (\sigma)
   (\omega-v k_x +i \epsilon ).
   \label{implicit2}
\end{equation}

The dispersion, $\omega(k_x)$, given by the location of the singularity of $G(\omega, k_x)$ is no longer monotonic as in the free case. The singularity occurs when the roots of  \eqref{implicit2} leave the real line. The dispersion can not be found analytically in general but for the two points where $\mathrm{d}\omega/\mathrm{d}k_x=0$ we have,
  \begin{equation}
  \begin{split}
\omega=\pm\lambda^2\frac{\sqrt{1-v^2}-\cosh ^{-1}\left(v^{-1}\right)}{4\pi(1-v^2)^{3/2}},\\
k_x=\pm\lambda^2\frac{\sqrt{1-v^2}-v^2\cosh ^{-1}\left(v^{-1}\right)}{4\pi v(1-v^2)^{3/2}}.
  \end{split}
  \end{equation}

\section{Integrals of Spectral Function}
\label{sec:integr-spectr-funct}

Several important observables like the density of states or the occupation number are defined by momentum space integrals of the spectral function 
$A(\omega, k)=-2\im G_R(\omega,k)$. Despite the fact that we have only an implicit expression for the Green's function \eqref{GreenImpl}, 
these integrals can be relatively easily evaluated by bringing the imaginary axis projection outside the integral and then changing integration variable to $\sigma$. We then do not have a closed form expression for the (now complex) contour of integration but the integrand is greatly simplified.
  
For a fixed $k_x$ we have $\omega$ as a closed form function of $\sigma$. Making this change of variable in integrals over $\omega$ gives the integrand
\begin{equation}
\int_C\mathrm{d}\omega A(\omega,k_x)=-2\im\left(\int_{\sigma(C)}\mathrm{d}\sigma \frac{\sinh (\sigma)}{v-\cosh (\sigma)}\right).\label{trick}
  \end{equation}
The curve of integration, $\sigma(C)$, is now defined through the implicit expression for $\sigma$ in  \eqref{implicit2}.

\begin{figure}
\includegraphics{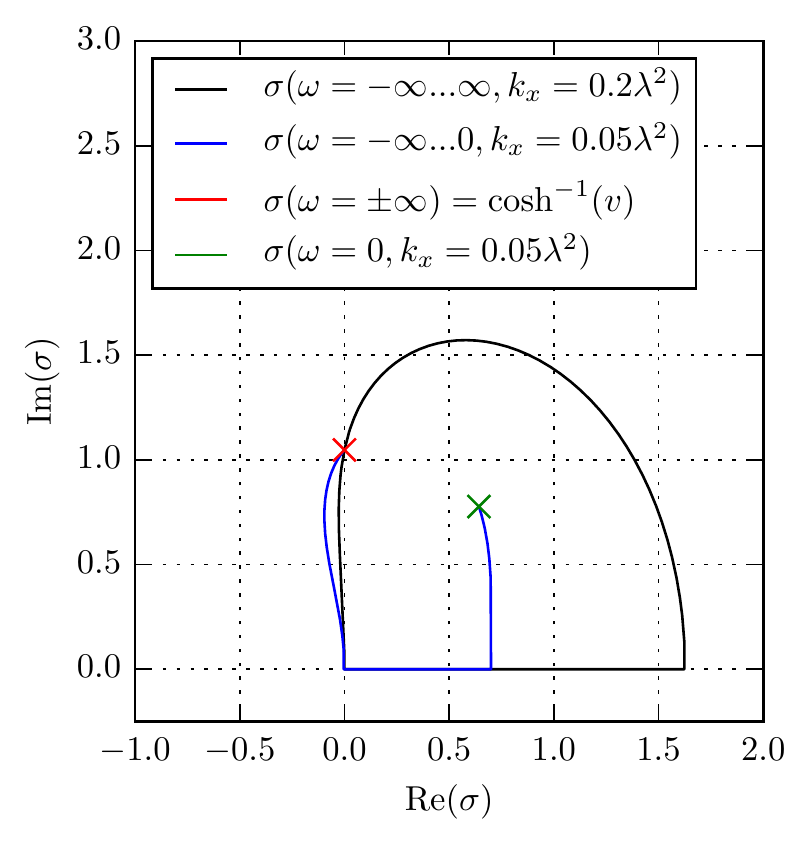}
\protect\caption{This figure shows a closed contour of integration for the sum rule and an open contour for calculating the occupation number integral. $v=.5$.}
\label{occupationIntegralFig}
\end{figure}

First of all we check that the sum rule $\int\mathrm{d}\omega A(\omega,k_x)=2\pi$ is satisfied. Taking the $\omega\rightarrow\pm\infty$ limits in  \eqref{implicit2} we see that $\sigma$ approaches $\cosh^{-1}(v)$ in both limits and the curve is thus closed. See Fig.~\ref{occupationIntegralFig}. To solve the integral we thus just have to figure out what poles are within the contour. It turns out that the single pole is the one at $\sigma=\cosh^{-1}(v)$, which is on the contour. This gives divergences but since the residue is real they are in the real part and do not matter for the spectral density. The contribution to the imaginary part is just $2\pi i$ times half the residue since the contour is smooth at the pole. The result of the integral is then $2\pi$ as expected, for all values of $k_x/\lambda^2$ and $v$.

The occupation number at zero temperature is given by
 \begin{equation}
\rho(k_x)=\int_{-\infty}^0\frac{\mathrm{d}\omega}{2\pi}A(\omega,k_x).
 \end{equation}
Since this contour is not closed we find a primitive function defined along the whole contour. The contribution from the point $\omega=0$ depends on $\sigma(\omega=0,k_x,v)$ so we can not get a closed form expression in this case. The contribution from $\omega\rightarrow-\infty$ now depends on the direction of the limit in the complex $\sigma$-plane since the point is only approached from one side. Summing the contributions from the two endpoints of the integral gives
 \begin{equation}
\rho(k_x)=\frac{1}{\pi}\arg\left[
\frac{  \cosh(\sigma(\omega=0, k_x/\lambda^2, v))-v }{ v  \cosh^{-1}(v)   -\sqrt{1 - v^2} (i - 4 \pi (1-v^2)k_x/\lambda^2)  }\right].
 \end{equation}
From this we see that in the region where $\sigma$ is real we actually have a closed form expression for the occupation number.

\begin{figure}
\includegraphics{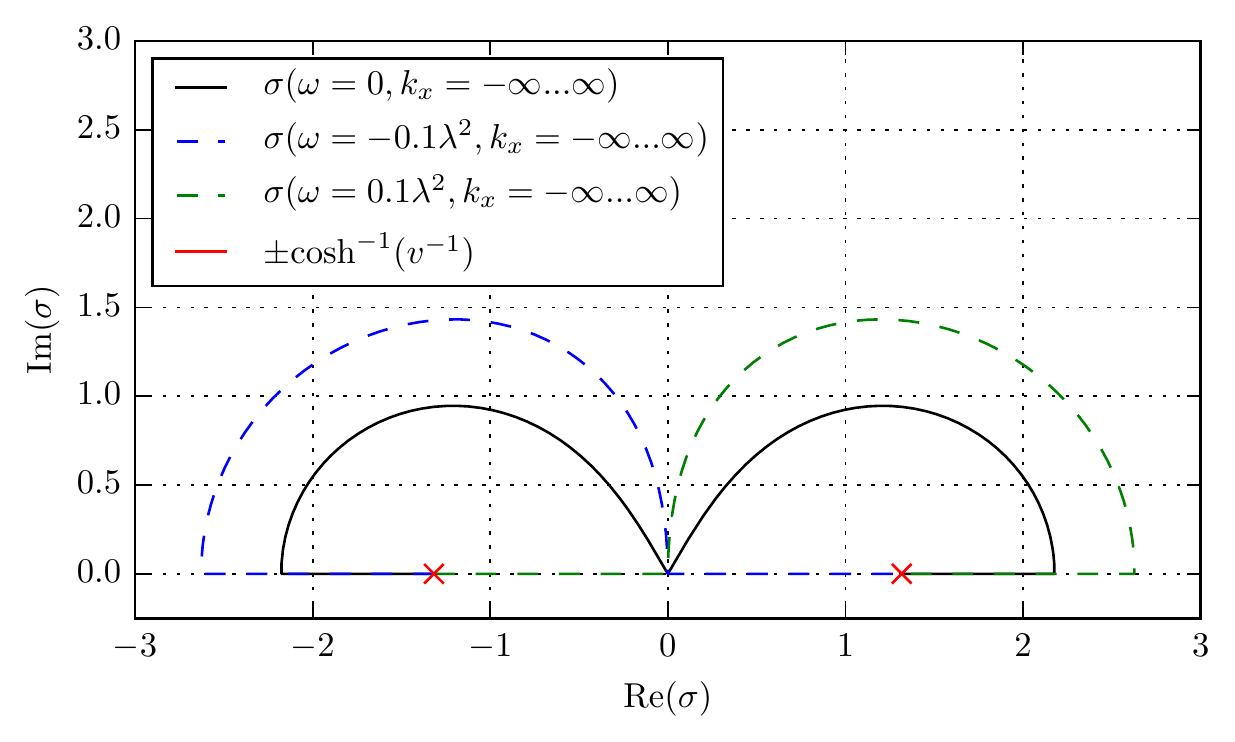}
\protect\caption{Integration contours for calculating density of states. The integrand and locations of the endpoints are independent of $\omega$ but since the integrand has poles at the endpoints the direction of approach matters. The poles are always approached along the real axis and this figure shows the three possible configurations. $v=.5$.}
\label{DoSIntegralFig}
\end{figure}

The density of states, $N(\omega)=\int\mathrm{d}k_x A(\omega, k_x)$ is similarly calculated by changing variables to $\sigma$. For any $\omega$ there is a $K_x$ such that $\sigma(k_x)$ is real for all $\lvert k_x\rvert>K_x$. The limits $k_x\rightarrow\pm\infty$ give $\sigma\rightarrow\pm\cosh^{-1}(1/v)$ and these are thus approached along the real line. Once again the integrand has poles (residue 1/v) at these points and since we are only interested in the imaginary part of the integral of the retarded Green's function we will only need to know the direction we approach these poles from. Finding a primitive function is again trivial and in the end the result only depends on the direction the poles are approached from. Since $\sigma$ is real in the limits, each pole is approached from either the left or the right. There are three different cases, for
 \begin{equation}
\omega<-\lambda^2\frac{\cosh ^{-1}\left(v^{-1}\right)-\sqrt{1-v^2}}{4\pi(1-v^2)^{3/2}}
\end{equation}
both poles are approached from the left. For 
\begin{equation}
\omega>\lambda^2\frac{\cosh ^{-1}\left(v^{-1}\right)-\sqrt{1-v^2}}{4\pi(1-v^2)^{3/2}}
\end{equation}
both poles are approached from the right and for $\omega$ between these two values the left pole is approached from the left and the right pole from the right. See Fig.~\ref{DoSIntegralFig}.
Taking these different limits of the primitive function gives
 \begin{equation}
N(\omega)=\frac{1}{v}\left[1+\theta(\lambda^2\frac{\cosh ^{-1}\left(v^{-1}\right)-\sqrt{1-v^2}}{4\pi(1-v^2)^{3/2}}-\lvert\omega\rvert)\right].
\end{equation}
The density of state takes two different values and we see that the $\omega$ where it changes are exactly the points where there are two instead of one solution in $k_x$ to the equation $G^{-1}(\omega, k_x)=0$.

\end{document}